\documentclass[preprint,aps,eqsecnum,draft]{revtex4}

\usepackage{amssymb,amsmath,amsfonts}
\usepackage{amsthm}

\def\eqref#1{(\ref{#1})}

\def\EQ{\begin{equation}}
\def\EQs{\begin{eqnarray}}
\def\endEQ{\end{equation}}
\def\endEQs{\end{eqnarray}}

\def\pos{{\vec r}}

\def\Im{{\rm Im}}
\def\Re{{\rm Re}}

\def\div{{\rm div}}
\def\tr{{\rm tr}}

\def\J{{\rm J}}

\def\eom/{equation of motion}
\def\esom/{equations of motion}

\newtheorem{theorem}{Theorem}{\bf}{\em}
\newtheorem{proposition}{Proposition}{\bf}{\em}
\newtheorem{lemma}{Lemma}{\bf}{\em}

\begin{document}


\title{Integrable generalizations of Schr\"odinger maps 
and Heisenberg spin models from Hamiltonian flows of curves and surfaces}

\author{Stephen C. Anco} 
\affiliation{ 
Department of Mathematics, Brock University, St. Catharines, ON Canada}
\email{sanco@brocku.ca}

\author{R. Myrzakulov}
\affiliation{ 
Department of General and Theoretical Physics, 
Eurasian National University, Astana, 010008, Kazakhstan}
\email{cnlpmyra@mail.ru}

\begin{abstract}
A moving frame formulation of 
non-stretching geometric curve flows in Euclidean space is used to derive 
a 1+1 dimensional hierarchy of integrable $SO(3)$-invariant vector models
containing the Heisenberg ferromagnetic spin model as well as 
a model given by a spin-vector version of the mKdV equation. 
These models describe a geometric realization of the NLS hierarchy of
soliton equations whose bi-Hamiltonian structure is shown to be encoded
in the Frenet equations of the moving frame. 
This derivation yields an explicit bi-Hamiltonian structure, 
recursion operator, and constants of motion for each model in the hierarchy. 
A generalization of these results to geometric surface flows is presented,
where the surfaces are non-stretching in one direction 
while stretching in all transverse directions. 
Through the Frenet equations of a moving frame, 
such surface flows are shown to encode a hierarchy of 
2+1 dimensional integrable $SO(3)$-invariant vector models,
along with their bi-Hamiltonian structure, recursion operator, 
and constants of motion, 
describing a geometric realization of 2+1 dimensional bi-Hamiltonian 
NLS and mKdV soliton equations. 
Based on the well-known equivalence between 
the Heisenberg model and the Schr\"odinger map equation in 1+1 dimensions, 
a geometrical formulation of these hierarchies of 1+1 and 2+1 vector models
is given in terms of dynamical maps into the 2-sphere. 
In particular, this formulation yields a new integrable generalization of 
the Schr\"odinger map equation in 2+1 dimensions
as well as a mKdV analog of this map equation corresponding to 
the mKdV spin model in 1+1 and 2+1 dimensions. 
\end{abstract}


\maketitle

\section{Introduction and Summary}

Spin systems are an important class of dynamical vector models 
from both physical and mathematical points of view. 
In physics such models describe nonlinear dynamics of magnetic materials,
while in mathematics they give rise to associated geometric flows of 
curves where the unit tangent vector along a curve is identified with 
a dynamical spin vector. 

A main example \cite{Lak} is the Heisenberg model 
for the dynamics of an isotropic ferromagnet spin system in 1+1 dimensions. 
The geometric curve flow described by this $SO(3)$-invariant model 
corresponds to the equations of motion of a non-stretching vortex filament 
in Euclidean space. 
Remarkably, 
the vortex filament equations are an integrable Hamiltonian system that is 
equivalent to the 1+1 dimensional focusing nonlinear Schr\"odinger equation 
(NLS) through a change of dynamical variables 
known as a Hasimoto transformation \cite{Has}. 

The vortex filament equations are one example 
in an infinite hierarchy of non-stretching geometric flows of space curves 
whose equations of motion have a well-understood integrability: e.g.
a Lax pair and an associated isospectral linear eigenvalue problem;
an infinite set of symmetries and constants of motion; 
and exact solutions with solitonic properties. 
This integrability structure turns out to have a simple geometric origin. 
In particular, 
all of these equations of motion are generated through a recursion operator 
that can be derived geometrically \cite{NakSegWad,DolSan}
from the Serret-Frenet structure equations 
given by a $SO(3)$ moving frame formulation 
for arbitrary non-stretching curve flows in Euclidean space,
with the components of the frame connection matrix 
providing the dynamical variables 
that appear in the equations of motion. 
More recently, these $SO(3)$ frame structure equations have been found to 
geometrically encode a pair of compatible Hamiltonian operators that
yield a concrete bi-Hamiltonian structure for the equations of motion 
of each integrable curve flow in the hierarchy \cite{MarSanWan}. 

The explicit bi-Hamiltonian form of the resulting equations of motion 
depends on a choice of the $SO(3)$ moving frame 
for the underlying space curve, 
which determines the form of the frame connection matrix 
and hence yields the dynamical variables in terms of the curve. 
In the case of the vortex filament equations,
the dynamical variables consist of the curvature invariant, $\kappa$,
and the torsion invariant, $\tau$, of the space curve,
corresponding to the choice of a classical Frenet frame \cite{Gug} given by 
the unit tangent vector, unit normal and bi-normal vectors, along the curve. 
Other geometrical choices of a moving frame can be made \cite{Anco06}, 
since there is a $SO(3)$ gauge freedom relating any two orthonormal frames
along an arbitrary curve in Euclidean space. 
In particular, the Hasimoto transformation arises geometrically 
as a gauge transformation from a Frenet frame to a parallel frame \cite{Bis},
where the frame vectors in the normal space of the curve are chosen 
such that their derivative with respect to the arclength $s$ along the curve 
lies in the tangent space of the curve. 
This choice of frame is unique up to rigid $SO(2)$ rotations 
acting on the normal vectors by the same angle at all points along the curve, 
while leaving invariant the tangent vector. 
The corresponding pair of dynamical variables 
(defined by the connection matrix of a parallel frame) 
are naturally equivalent to a single complex-valued variable 
$u=\kappa\exp(i\int \tau ds)$
that is determined by the curve only up to constant phase rotations 
$u\rightarrow e^{i\phi} u$ 
(where $\phi$ is independent of arclength $s$). 
This dynamical variable $u$ thus has the geometrical meaning \cite{Anco08} 
of a $U(1)\simeq SO(2)$ covariant of the space curve. 
Importantly, 
the resulting Hamiltonian structure for the equations of motion 
looks simplest in terms of the covariant $u$, 
which directly incorporates the Hasimoto transformation, 
rather than using the classical invariants $\kappa$ and $\tau$. 

The purpose of the present paper will be to give 
some new applications of these ideas to the study of 
integrable vector models in 1+1 and 2+1 dimensions. 

Firstly, 
from the hierarchy of non-stretching geometric space curve flows 
that contains the vortex filament equations,
we derive the complete hierarchy of corresponding 
integrable $SO(3)$-invariant vector models in 1+1 dimensions,
along with their bi-Hamiltonian integrability structure in explicit form. 
In addition to the Heisenberg model, this hierarchy will be seen to contain
a model that describes a spin-vector version of the mKdV equation. 
Our results provide a new derivation of the Hamiltonian structure,
recursion operator, and constants of motion for these models. 

Secondly, 
we extend the derivation to a geometrically analogous class of surface flows 
where the surface is non-stretching in one coordinate direction 
while stretching in all transverse directions. 
Such surfaces arise in a natural fashion from a spatial Hamiltonian flow of
non-stretching space curves. 
This generalization will be shown to give rise to a class of 
2+1 dimensional NLS and mKdV soliton equations 
with an explicit bi-Hamiltonian structure, 
yielding a hierarchy of integrable $SO(3)$-invariant vector models 
in 2+1 dimensions.
In particular, this hierarchy includes 2+1 generalizations of
the Heisenberg spin model and the mKdV spin model, 
which were found in earlier work by one of us 
\cite{Myr,Myr1,Myr2,Myr3,Myrbook}. 
Our derivation here, in contrast, 
yields the explicit bi-Hamiltonian structure, recursion operator, 
and constants of motion, which are new results for these models. 
We also write out the corresponding surface flows explicitly in terms of
geometric variables given by \cite{Gug} 
the geodesic and normal curvatures and the relative torsion of
the non-stretching coordinate lines on the surface. 
The surface flow arising from the 2+1 integrable Heisenberg model 
will be seen to describe a sheet of non-stretching vortex filaments
in Euclidean space. 

Lastly, 
we also derive an interesting geometric formulation of these results 
by viewing the spin vector as a dynamical map 
into the 2-sphere in Euclidean space. 
This formulation is based on the well-known geometrical equivalence between
the Heisenberg model and the Schr\"odinger map equation 
in 1+1 dimensions \cite{Sha}. 
When applied to the 1+1 and 2+1 dimensional hierarchies 
of $SO(3)$-invariant vector models, 
our derivation yields a new integrable generalization of 
the Schr\"odinger map equation in 2+1 dimensions
as well as a new mKdV analog of this map equation corresponding to 
the mKdV spin-vector model in 1+1 and 2+1 dimensions. 

The rest of the paper is organized as follows. 
In section~II, 
we review from a unified point of view the mathematical relationships amongst 
1+1 dimensional vector models, 
dynamical maps into the 2-sphere, 
non-stretching curve flows in Euclidean space, 
Frenet and parallel frames, and the Hasimoto transformation. 
In section~III, 
we derive the NLS hierarchy of soliton equations in terms of 
the geometrical covariant $u$ given by the Frenet equations of 
a moving parallel frame for non-stretching space curve flows. 
This approach directly yields the explicit bi-Hamiltonian structure of 
these soliton equations, including a formula for the Hamiltonians. 
As examples, the parallel-frame Frenet equations are used to show, 
firstly, how the NLS equation itself corresponds geometrically to 
the Heisenberg spin model and the Schr\"odinger map equation; 
and secondly, how the mKdV spin model and the mKdV map equation 
arise geometrically from the next soliton equation in the NLS hierarchy. 

Section~IV contains several main results. 
We work out the equations of motion for the space curves 
corresponding to the NLS hierarchy
and write down the induced flows on the curvature and torsion invariants
$\kappa,\tau$. 
Next we derive the resulting geometrical hierarchies of
vector models and dynamical map equations, along with their
bi-Hamiltonian structure, recursion operators, and constants of motion. 
This new derivation involves only the parallel-frame Frenet equations
plus the bi-Hamiltonian structure of the NLS hierarchy. 
The explicit bi-Hamiltonian form of the Schr\"odinger map equation
and Heisenberg model, including a geometric expression for the Hamiltonians,
are presented as examples. 

In section~V, 
we consider surfaces generated by a spatial Hamiltonian flow of curves
with a parallel framing in Euclidean space. 
The underlying Hamiltonian structure is shown to arise naturally 
from the Frenet equations of the induced frame along the surface. 
This formulation is then used in section~VI 
to study surface flows expressed in terms of the covariant variable $u$ 
geometrically associated with the non-stretching space curves 
that foliate the surface, 
where the surface is stretching in all directions transverse to these curves.
We show that the bi-Hamiltonian structure for 1+1 flows on $u$ 
has a natural extension to 2+1 flows based on the observation 
that the Hamiltonian operators involve only the coordinate 
in the non-stretching direction on the surface. 
This leads to a hierarchy of 2+1 flows on $u$, 
with the starting flow given geometrically by translations 
in the coordinate in the transverse direction,
which yields a 2+1 generalization of the NLS hierarchy. 

The final two sections of the paper contain our main new results. 
In section~VII, 
we use the surface Frenet equations to derive the complete hierarchies of
integrable 2+1 vector models and dynamical maps arising from 
the 2+1 generalization of the NLS hierarchy. 
The derivation yields the explicit bi-Hamiltonian structure of 
these two hierarchies, in addition to their respective 
recursion operators and constants of motion. 
As examples, 
the integrable generalizations of the Heisenberg model 
and the mKdV spin model in 2+1 dimensions are written down in detail,
as well as the corresponding new 2+1 dimensional integrable generalizations 
of the Schr\"odinger map equation and mKdV map equation. 
In section~VIII, 
we work out the equations of motion for the surface flows that correspond
to the previous hierarchies. 
These equations are obtained by means of 
a different framing defined in a purely geometrical fashion 
by the non-stretching coordinate direction on the surface
and the orthogonal direction of the surface normal in Euclidean space. 
We also discuss aspects of both the intrinsic and extrinsic geometry of
the resulting surface motions. 
In particular, we obtain a recursion operator, constants of motion, 
and explicit evolution equations formulated in terms of geometric variables
given by the geodesic curvature, normal curvature, and relative torsion 
of the non-stretching coordinate lines on the surface. 

Some concluding remarks on future extensions of this work are given 
in section~IX.

\section{Vector models and space curve flows}

We start from an arbitrary $SO(3)$ vector model in 1+1 dimensions, 
\EQ
S_t=f(S,S_x,S_{xx}\ldots),\quad
|S|=1
\label{O3model}
\endEQ
where
$S(t,x)=(S_{1}, S_{2}, S_{3})$ is a dynamical unit vector in Euclidean space, 
$f$ is a vector function $\perp$ $S$,
and $x$ belongs to some one-dimensional domain $C$. 
A running example will be the Heisenberg spin model
\EQ
S_t=S\wedge S_{xx}=(S\wedge S_x)_x
\label{heisenbergmodel}
\endEQ
with $C$ being $\mathbb{R}$ or $\rm S^1$.

There are two different ways to associate a curve flow 
to equation \eqref{O3model}. 
One formulation consists of intrinsically identifying $S$ with
a map $\gamma$  into the unit sphere ${\rm S}^2\subset \mathbb{R}^3$.
Then $S_t$ and $S_x$ correspond to $\gamma_t$ and $\gamma_x$;
$\partial_x+S(S_x\cdot )$ corresponds to the covariant derivative
$\nabla_x$ on the sphere
with respect to the tangent direction $\gamma_x$; 
and $S\wedge$ corresponds to the Hodge dual $*=J$
(i.e. a complex structure on the sphere).
Under these identifications, 
each vector model \eqref{O3model} describes
a curve flow
\EQ
\gamma_t=F(\gamma_x,\nabla_x\gamma_x, \ldots)
\endEQ
for $\gamma(t,x)$ on ${\rm S}^2$.
Ex. the Heisenberg model \eqref{heisenbergmodel} corresponds to
\EQ
\gamma_t=J\nabla_x\gamma_x
\label{schrodingermap}
\endEQ
which is the Schr\"odinger map equation on ${\rm S}^2$.

Alternatively, in an extrinsic formulation, 
$S$ can be identified with the unit tangent vector $T$
along a non-stretching space curve given by a position vector $\pos$ 
in Euclidean space, 
\EQ
S=T=\pos_x
\endEQ
where $x$ is the arclength along the curve $\pos(x)$.
Then the equation of motion of $\pos$ is
\EQ
\pos_{tx}=f(\pos_x,\pos_{xx},\ldots),\quad
|\pos_x|=1 ,
\endEQ
or equivalently
\EQ
\pos_{t}=\int^x f(\pos_x,\pos_{xx},\ldots)dx,\quad
|\pos_x|=1 ,
\label{curveflow}
\endEQ
under which the arclength of the curve is preserved, i.e.
$\int_C |\pos_x| dx =\ell$ is a constant of the motion. 
Ex. the Heisenberg model \eqref{heisenbergmodel} corresponds to 
\EQ\label{filamentmotion}
\pos_{t}=\pos_x\wedge \pos_{xx}
\endEQ
with $|\pos_{x}|=1$.
This is the equation of motion of a non-stretching vortex filament
studied by Hasimoto \cite{Has}. 

To proceed we first introduce a Frenet frame $\mathbf{E}$ along $\pos(x)$.
It is expressed in matrix column notation by
\EQ
\mathbf{E}=
\begin{pmatrix}
T\\ N\\ B
\end{pmatrix}
\label{frenet}
\endEQ
where
\EQ
T=\pos_x,\quad
N=|T_x|^{-1} T_x=|\pos_{xx}|^{-1}\pos_{xx},\quad
B=T\wedge N=|\pos_{xx}|^{-1}\pos_x\wedge\pos_{xx}.
\label{TNB}
\endEQ
Here $N$ is the unit normal and $B$ is the unit bi-normal 
of the space curve $\pos(x)$.
Note we have the relations 
\EQ
S=T,\quad
S_x=\kappa N,\quad
S\wedge S_x=\kappa B,
\label{Sframe}
\endEQ
where
\EQ
\kappa=T_x\cdot N=|S_x|
\endEQ
is the curvature of $\pos(x)$,
and
\EQ
\tau=N_x\cdot B=|S_x|^{-2} S_{xx}\cdot(S\wedge S_x)
\endEQ
is the torsion of $\pos(x)$.
The Serret-Frenet equations of this frame \eqref{frenet} are given by 
\EQ
\mathbf{E}_x=\mathbf{KE}
\label{xfrenet}
\endEQ
with
\EQ
\mathbf{K}=
\left(\begin{array}{ccc} 
0 & \kappa & 0\\ -\kappa & 0 & \tau \\ 0 & -\tau 
& 0
\end{array}\right)
\in \mathfrak{so}(3).
\label{K}
\endEQ

From the equation of motion \eqref{O3model} for $S$
we obtain the frame evolution equation
\EQ
\mathbf{E}_t=\mathbf{AE} ,\quad
\mathbf{A}=
\left(\begin{array}{ccc} 0 & a_2 & a_3 \\ -a_2 & 0 & a_1 \\ -a_3 & -a_1
& 0\end{array}\right)\in \mathfrak{so}(3)
\endEQ
where 
\EQs
&&
a_1=f_x\cdot B/|S_x| =f_x\cdot (S\wedge S_x)/|S_x|^2 ,\nonumber\\
&&
a_2=f\cdot N =f\cdot S_x/|S_x| ,\nonumber\\
&&
a_3=f\cdot B =f\cdot (S\wedge S_x)/|S_x| ,\nonumber
\endEQs
are determined by taking the $t$-derivative of \eqref{Sframe}
and substituting \eqref{O3model}, followed by applying respective projections 
orthogonal to $T,N,B$. 

This evolution of the frame $\mathbf{E}$ induces evolution equations
for $\kappa$ and $\tau$ through the zero-curvature relation
$\mathbf{K}_t=\mathbf{A}_x+[\mathbf{A},\mathbf{K}]$.
Ex. the Heisenberg model \eqref{heisenbergmodel} gives 
the vortex filament equations in terms of the curvature and torsion
\cite{Lak}:
\EQs
&&
\kappa_t
=-\kappa \tau_x-2\kappa_{x}\tau
=-\frac{(\kappa^2\tau)_x}{\kappa} , \nonumber\\
&&
\tau_t
= \frac{\kappa_{xxx}}{\kappa} - \frac{\kappa_{xx}\kappa_x}{\kappa^2}
-2\tau\tau_x+\kappa\kappa_{x} 
= (\frac{\kappa_{xx}}{\kappa}-\tau^2+\frac{1}{2}\kappa^2)_{x} .\nonumber\\
\label{vfe}
\endEQs

Next we perform a $SO(2)$ gauge transformation on the normal vectors
in the Frenet frame \eqref{frenet}:
\EQ
\widetilde{E}_1=E_1=T, \quad
\widetilde{E}_2=E_2\cos\theta+E_3\sin\theta,\quad
\widetilde{E}_3=-E_2\sin\theta+E_3\cos\theta
\label{gaugetransform}
\endEQ
with the rotation angle $\theta$ defined by
\EQ
\theta_x=-\tau
\label{parallelgauge}
\endEQ
so thus
\EQs
&&
\widetilde{E}_{1x}=
\kappa\cos\theta \widetilde{E}_2
-\kappa\sin\theta\widetilde{E}_3 \quad \perp\ T , 
\label{E1x}
\\
&&
\widetilde{E}_{2x}=-\kappa\cos\theta \widetilde{E}_1 \quad \parallel\ T, 
\quad
\widetilde{E}_{3x}=\kappa\sin\theta \widetilde{E}_1 \quad \parallel\ T.
\label{E23x}
\endEQs
This is called a {\em parallel framing} \cite{Bis} 
of the space curve $\pos(x)$.
The frame vectors \eqref{gaugetransform} are characterized by 
the geometrical property that along $\pos(x)$ 
their derivatives lie completely in the normal space \eqref{E1x}
or in the tangent space \eqref{E23x}. 
Such a frame is unique up to a rigid ($x$-independent) rotation 
\EQ\label{U1group}
\theta\rightarrow \theta +\phi,\quad \phi=\text{const.}
\endEQ 
acting on the pair of normal vectors.

In matrix notation
the Serret-Frenet equations of a parallel frame 
are given by
\EQ\label{xparallel}
\widetilde{\mathbf{E}}_x=\mathbf{U}\widetilde{\mathbf{E}} 
\endEQ
with
\EQ
\widetilde{\mathbf{E}}=
\left(\begin{array}{ccc}
T \\ \cos\theta N+\sin\theta B \\ -\sin\theta N+\cos\theta B
\end{array}\right)
,\quad
\mathbf{U}=\left(\begin{array}{ccc}
0 & u_2& u_3\\ -u_2 & 0 & u_1\\ -u_3 & -u_1 & 0
\end{array}\right)\in \mathfrak{so}(3)
\label{parallel}
\endEQ
where
\EQ
u_1=0,\quad
u_2=\kappa\cos\theta=\kappa\cos(\textstyle\int \tau dx),\quad
u_3=-\kappa\sin\theta=\kappa\sin(\textstyle\int \tau dx)
\label{parallelmatrix}
\endEQ
are the components of the principal normal $T_x$ of $\pos(x)$.
The evolution of this frame
\EQ
\widetilde{\mathbf{E}}_t=\mathbf{W}\widetilde{\mathbf{E}}
\label{Et}
\endEQ
is described by the matrix
\EQ
\mathbf{W}=
\left(\begin{array}{ccc} 0 & \varpi_2& \varpi_3\\
-\varpi_2 & 0 & \varpi_1\\ -\varpi_3 &
-\varpi_1 & 0 \end{array}\right)\in \mathfrak{so}(3)
\label{W}
\endEQ
which is related to $\mathbf{U}$
through the zero-curvature equation
\EQ
\mathbf{U}_t-\mathbf{W}_x+[\mathbf{U},\mathbf{W}]=0.
\label{UWeq}
\endEQ
Note $\mathbf{W}$ can be determined directly from the model \eqref{O3model}
via the relations \eqref{frenet} and \eqref{Sframe}. 

It now becomes convenient to work in terms of a complex variable
formalism
\EQs
&&
\varpi=\varpi_2+i\varpi_3, 
\label{w}\\
&&
u=u_2+iu_3=\kappa e^{-i\theta}=\kappa \exp(i\textstyle\int\tau dx), 
\label{u}
\endEQs
encoding the well-known Hasimoto transformation \cite{Has}. 
Ex. in the Heisenberg model \eqref{heisenbergmodel}, 
the vortex filament equations on $\kappa$ and $\tau$ 
transform into the NLS equation on $u$:
\EQ
-iu_t=u_{xx}+\frac{1}{2}|u|^2u. 
\label{nls}
\endEQ
Thus, 
Hasimoto's transformation has the geometrical interpretation \cite{DolSan}
of a $SO(2)$ gauge transformation on the normal frame of the curve $\pos(x)$,
relating a Frenet frame to a parallel frame.

{\bf Remark:}
Since the form \eqref{parallelmatrix} of a parallel frame is preserved
by $SO(2)$ rotations \eqref{U1group}, 
the complex scalar variable \eqref{u} given by the Hasimoto transformation
is uniquely determined by the curve $\pos(x)$ up to rigid phase rotations
$u\rightarrow e^{-i\phi} u$, depending on an arbitrary constant $\phi$. 
Therefore, $u$ has the geometrical meaning of a {\em covariant}
of the curve \cite{Anco08} relative to the group $SO(2)\simeq U(1)$,
while $|u|=\kappa$ and $(\arg u)_x =\tau$ are invariants of the curve.

\section{Bi-Hamiltonian flows and operators}

For a general vector model \eqref{O3model} 
the zero-curvature equation \eqref{UWeq}
gives an evolution equation on $u$, 
\EQ
u_t=\varpi_x-i\varpi_{1} u,
\label{ueq}
\endEQ
plus an auxiliary equation relating $\varpi_1$ to $u$, 
\EQ
\varpi_{1x}=\Im(\bar\varpi u) .
\label{w1eq}
\endEQ
From \eqref{w1eq} we can eliminate
$\varpi_1=D^{-1}_x\Im(\bar\varpi u)$ in terms of $u$ and $\varpi$,
and then we see \eqref{ueq} yields 
\EQ
u_t=D_x\varpi -iuD^{-1}_x\Im(\bar\varpi u)=\mathcal{H}(\varpi)
\label{uflow}
\endEQ
where $\varpi$ is determined from \eqref{O3model}
via the frame evolution equation \eqref{Et}. 

\begin{proposition}\label{uHamstructure}
\EQ
\mathcal{H}=D_x -iu D_x^{-1}\Im(u\mathcal{C})
\label{Hop}
\endEQ
is a Hamiltonian operator with respect to the flow variable $u(t,x)$,
whence the evolution equation \eqref{uflow} has a Hamiltonian structure
\EQ
u_t=\mathcal{H}(\delta \mathfrak{H}/\delta \bar u)
\label{uHamflow}
\endEQ
iff
\EQ
\varpi=\delta H/\delta \bar u
\label{Hamgrad}
\endEQ
holds for some Hamiltonian 
\EQ
\mathfrak{H}=\int_C H(x,u,\bar u,u_x,\bar u_x,u_{xx},\bar u_{xx},\ldots) dx .
\label{Hamfunct}
\endEQ
\end{proposition}

Here $\mathcal C$ is the complex conjugation operator,
and $C=\mathbb{R}$ or $\rm S^1$ is the domain of $x$. 
In the present setting,
an operator $\mathcal{D}$ is {\em Hamiltonian} 
if it defines an associated Poisson bracket
\EQ
\{\mathfrak{H},\mathfrak{G}\} = 
\int_C\Re( \mathcal{D}(\delta\mathfrak{H}/\delta\bar u) \delta\mathfrak{G}/\delta u )dx
\label{PB}
\endEQ
obeying skew-symmetry 
$\{\mathfrak{H},\mathfrak{G}\} =-\{\mathfrak{G},\mathfrak{H}\}$
and the Jacobi identity
$\{\mathfrak{F},\{\mathfrak{H},\mathfrak{G}\}\} +\text{cyclic}=0$, 
for all real-valued functionals $\mathfrak{F}$, $\mathfrak{G}$, $\mathfrak{H}$
on the $x$-jet space of the flow variable $u$. 

\begin{proposition}\label{Hamops}
(i) 
The Hamiltonian operator $\mathcal{H}$ is invariant with respect to 
$U(1)$ phase rotations 
$e^{i\lambda}\mathcal{H} e^{-i\lambda} = 
\mathcal{H}|_{u\rightarrow e^{i\lambda}u}$. 
(ii) 
A second Hamiltonian operator is given by 
\EQ
\mathcal{I}=-i 
\endEQ
which is similarly $U(1)$-invariant, 
$e^{i\lambda}\mathcal{I} e^{-i\lambda} = \mathcal{I}$. 
(iii) 
The operators $\mathcal{H}$ and $\mathcal{I}$ 
are a compatible Hamiltonian pair
(i.e. every linear combination is again a Hamiltonian operator),
and their compositions define $U(1)$-invariant hereditary recursion operators
\EQ
\mathcal{R}=\mathcal{H}\mathcal{I}^{-1} 
=i(D_x +u D_x^{-1}\Re(u\mathcal{C})) ,\quad
\mathcal{R}^*=\mathcal{I}^{-1}\mathcal{H}
=iD_x -u D_x^{-1}\Re(iu\mathcal{C}) .
\label{Rops}
\endEQ
(iv)
Composition of $\mathcal{R}$ and $\mathcal{H}$ yields a third
$U(1)$-invariant Hamiltonian operator
\EQs
\mathcal{E} =\mathcal{R}\mathcal{H}
&=& 
iD_x^2 +D_x(u D_x^{-1}\Im(u\mathcal{C})) + iu D_x^{-1}\Re(u D_x\mathcal{C})
\\
&=&
iD_x^2 +i|u|^2 +u_x D_x^{-1}\Im(u\mathcal{C}) -iu D_x^{-1}\Re(u_x\mathcal{C})
\nonumber
\endEQs
satisfying 
$e^{i\lambda}\mathcal{E} e^{-i\lambda} = 
\mathcal{E}|_{u\rightarrow e^{i\lambda}u}$. 
In particular, $\mathcal{E}$, $\mathcal{H}$, $\mathcal{I}$
form a compatible Hamiltonian triple.
\end{proposition}

These Propositions are a special case of group-invariant
bi-Hamiltonian operators derived from non-stretching curve flows
in constant-curvature spaces and general symmetric spaces 
in recent work \cite{SanWan,Anco06b,Anco07,Anco08}. 
Moreover, in the present complex variable formalism,
Proposition~\ref{Hamops} provides a substantial simplification of
some main results in \cite{MarSanWan} on Hamiltonian operators
connected with non-stretching curve flows in Euclidean space. 

Because phase rotation on $u$ is a symmetry of both 
$\mathcal{H}$ and $\mathcal{I}$,
the recursion operator $\mathcal{R}$
generates a hierarchy of commuting Hamiltonian vector fields given by
\EQ
i\varpi^{(n)} \partial/\partial u =\mathcal{R}^n(iu) \partial/\partial u ,
\quad n=0,1,2,\ldots
\label{nthHamvectorfield}
\endEQ
where
\EQ
\varpi^{(n)}=\delta H^{(n)}/\delta \bar u=\mathcal{R}^{*n}(u) ,
\quad n=0,1,2,\ldots
\label{nthHamgrad}
\endEQ
are Hamiltonian derivatives, 
starting with
\EQ
\varpi^{(0)}=u,\quad
H^{(0)}={\bar u}u=|u|^2
\endEQ
which corresponds to phase-rotation 
$iu\partial/\partial u$. 
Next in the hierarchy comes
\EQ
\varpi^{(1)}=iu_x,\quad
H^{(1)}=\frac{i}{2}({\bar u}u_x-u{\bar u}_x)=\Im(\bar u_x u) , 
\endEQ
followed by
\EQ
\varpi^{(2)}=-(u_{xx}+\frac{1}{2}|u|^2u), \quad
H^{(2)}=|u_x|^2-\frac{1}{4}|u|^4 , 
\endEQ
corresponding to respective Hamiltonian vector fields 
$-u_x\partial/\partial u$
which is $x$-translation and 
$-i(u_{xx}+\frac{1}{2}|u|^2u)\partial/\partial u$ 
which is of NLS form.

Through Propositions~\ref{uHamstructure} and~\ref{Hamops}, 
this hierarchy produces integrable evolution equations on $u(t,x)$ 
with a tri-Hamiltonian structure. 
An explicit formulation of this result has not appeared previously 
in the literature. 

\begin{theorem}\label{uhierarchy}
There is a hierarchy of integrable bi-Hamiltonian flows on $u(t,x)$
given by
\EQ
u_t=\mathcal{H}(\delta \mathfrak{H}^{(n)}/\delta\bar u) 
=\mathcal{I}(\delta \mathfrak{H}^{(n+1)}/\delta \bar u), \quad 
n=0,1,2,\ldots 
\label{ut}
\endEQ
(called the {\em $+n$ flow})
in terms of Hamiltonians $\mathfrak{H}^{(n)}=\int_C H^{(n)} dx$
where 
\EQ
H^{(n)}=
\frac{2}{1+n} D^{-1}_x\Im( \bar u (i\mathcal{H})^{n+1}u )
\quad n=0,1,2,\ldots
\label{Ham}
\endEQ
are local Hamiltonian densities. 
Moreover, all the flows for $n\neq 0$ have a tri-Hamiltonian structure
\EQ
u_t=\mathcal{E}(\delta \mathfrak{H}^{(n-1)}/\delta\bar u) , \quad 
n=1,2,\ldots\quad . 
\label{triHam}
\endEQ
\end{theorem}

{\bf Remarks:}
Each flow $n=0,+1,+2,\ldots$ in the hierarchy
is $U(1)$-invariant under the phase rotation 
$u\rightarrow e^{i\lambda}u$
and has scaling weight $t\rightarrow \lambda^{1+n}t$
under the NLS scaling symmetry
$x\rightarrow\lambda x$, $u\rightarrow \lambda^{-1}u$,
where the scaling weight of $H^{(n)}$ is $-2-n$.
Additionally, these flows on $u(t,x)$ each admit constants of motion
(under suitable boundary conditions)
\EQ
D_t\int_C |u|^2 dx =0 ,\quad
D_t\int_C i{\bar u} u_x\, dx =0 ,\quad
D_t\int_C |u_x|^2 -\frac{1}{4} |u|^4 dx =0 ,\quad\ldots
\endEQ
and symmetries
\EQ
-u_x\partial/\partial u, \quad
-i(u_{xx}+\frac{1}{2}|u|^2u)\partial/\partial u ,\quad
(u_{xxx}+\frac{3}{2}|u|^2 u) \partial/\partial u ,\quad\ldots
\endEQ
respectively comprising 
all of the Hamiltonians \eqref{Ham} in the hierarchy
and all of the corresponding Hamiltonian vector fields \eqref{nthHamvectorfield}.

At the bottom of the hierarchy,
the $0$ flow is given by a linear traveling wave equation $u_t=u_x$,
and next the $+1$ flow produces the NLS equation \eqref{nls}. 
The $+2$ flow yields the complex mKdV equation
\EQ
-u_t=u_{xxx}+\frac{3}{2}|u|^2u 
\label{mkdv}
\endEQ
which corresponds to an mKdV analog of the vortex filament equations,
\EQs
-\kappa_t 
&=& 
(\kappa_{xx}+\frac{1}{2}\kappa^3)_x 
-\frac{3}{2} \frac{(\tau^2\kappa^2)_x}{\kappa} 
= \kappa_{xxx}+\frac{3}{2}(\kappa^2 -2\tau^2)\kappa_x -3 \kappa\tau\tau_x ,
\label{highervfe}\\
-\tau_t 
&=&
(\tau_{xx}+ 3\frac{(\tau\kappa_x)_x}{\kappa} 
+\frac{3}{2}\tau\kappa^2 -\tau^3)_x
\label{highervfe'}\\\nonumber
&=& 
\tau_{xxx} + 3\frac{\tau_{xx}\kappa_x}{\kappa}
+\tau_x( 6\frac{\kappa_{xx}}{\kappa} - 3\frac{\kappa_x^2}{\kappa^2} -3\tau^2 
+\frac{3}{2}\kappa^2 )
+\tau( 3\frac{\kappa_{xxx}}{\kappa} - 3\frac{\kappa_{xx}\kappa_x}{\kappa^2} 
+ 3\kappa\kappa_x ) ,
\endEQs
as obtained through the Hasimoto transformation $u=\kappa \exp(-i\theta)$. 

The evolution equations describing the $0,+1,+2,\ldots$ flows on $u$
each arise from geometric space curve flows 
corresponding to $SO(3)$-invariant vector models \eqref{O3model}.
To make this correspondence explicit, 
it is convenient to introduce a complex frame notation
\EQ
E^\parallel=\widetilde{E}_1 =T, \quad 
E^\perp=\widetilde{E}_2+i\widetilde{E}_3 = e^{-i\theta}(N+iB)
\label{complexE}
\endEQ
satisfying
\EQ
E^\parallel\wedge E^\perp= 
-iE^\perp=\widetilde{E}_3-i\widetilde{E}_2= e^{-i\theta}(B-iN)  
\label{iE}
\endEQ
and 
\EQ
E^\parallel\cdot E^\parallel =1, \quad
E^\perp \cdot \bar E^\perp= 2, \quad
E^\parallel\cdot E^\perp= 0=E^\perp \cdot E^\perp .
\label{complexEnorms}
\endEQ
The Frenet equations \eqref{xfrenet} become
\EQ
E^\parallel_{x}=\Re(\bar u E^\perp) ,\quad
E^\perp_{x}=-uE^\parallel , 
\label{complexEx}
\endEQ
while from \eqref{W}, \eqref{w}, \eqref{w1eq}, 
the evolution of the frame is given by the equations
\EQ
E^\parallel_{t}=
\Re(\bar\varpi E^\perp) ,\quad
E^\perp_{t}=
iD_x^{-1}\Im(\varpi\bar u) E^\perp-\varpi E^\parallel . 
\label{complexEt}
\endEQ
Then any flow belonging to the general class
\EQ
\varpi=\varpi(u,\bar u,u_x,\bar u_x,u_{xx},\bar u_{xx},\ldots) 
\label{flows}
\endEQ
will determine a vector model \eqref{O3model} via 
\EQ
S=E^\parallel,\quad  f=\Re(\bar\varpi E^\perp),
\label{Sfrels}
\endEQ
where $f$ is expressed in terms of $S$, $S_x$, $S_{xx}$, etc.
through the Frenet equations \eqref{xparallel}--\eqref{parallel}. 

{\bf Ex. 1}:
The $+1$ flow $\varpi=iu_x$ yields
\EQ
E^\parallel_t=-\Re(i{\bar u}_x E^\perp) . 
\endEQ
By rewriting
$$
{\bar u}_x E^\perp=( {\bar u}E^\perp)_x + {\bar u}u E^\parallel
$$
we obtain 
\EQs
&
\Re(i{\bar u}E^\perp)
= - E^\parallel\wedge \Re({\bar u}E^\perp) 
=- E^\parallel\wedge E^\parallel_{x},
\nonumber\\
&
\Re(i{\bar u}u E^\parallel) =\Re(i|u|^2)E^\parallel =0, 
\nonumber
\endEQs
and hence
\EQ
E^\parallel_t=(E^\parallel\wedge E^\parallel_x)_x . 
\endEQ
The identifications \eqref{Sfrels} then directly give 
the $SO(3)$ Heisenberg model \eqref{heisenbergmodel},
which corresponds to the non-stretching space curve flow \eqref{vfe} 
or equivalently 
\EQ 
\pos_t = \kappa B ,\quad |\pos_x|=1
\endEQ
expressed as a geometric flow. 

{\bf Ex. 2}:
The $+2$ flow
$\varpi=-(u_{xx}+\frac{1}{2}|u|^2u)$ yields
\EQ
-E^\parallel_{t}=
\Re(({\bar u}_{xx}+\frac{1}{2}|u|^2{\bar u})E^\perp) 
=\Re({\bar u}_{xx}E^\perp)+\frac{1}{2}|u|^2E^\parallel_{x}.
\endEQ
Here we can rewrite the first term as
$$
\Re(\bar u_{xx}E^{\perp})
=\Re(\bar u E^\perp)_{xx}
-\Re(\bar u E^\perp_x)_x
-\Re(\bar u_x E^\perp_x)
=E^\parallel_{xxx}+(|u|^2E^\parallel)_x+\frac{1}{2}({\bar u}u)_x E^\parallel,
$$
with $|u|^2=|E^\parallel_{x}|^2$,
and thus
\EQ
-E^\parallel_{t}=
E^\parallel_{xxx}+\frac{3}{2}(|E^\parallel_x|^2E^\parallel)_x.
\endEQ
Hence,
$E^\parallel=S$ gives
\EQ
-S_t=S_{xxx}+\frac{3}{2}(|S_x|^2S)_x
\label{mkdvmodel}
\endEQ
which can be viewed as an $SO(3)$ mKdV model.
The corresponding non-stretching space curve flow looks like
\EQ
-\pos_t = \pos_{xxx} +\frac{3}{2}|\pos_{xx}|^2 \pos_x ,\quad
|\pos_x|=1 .
\label{mkdvspacecurve}
\endEQ
This describes a geometric flow \cite{LangerPerline}
\EQ
-\pos_t = \frac{1}{2}\kappa^2 T +\kappa_x N +\kappa\tau B ,\quad |\pos_x|=1
\endEQ
which is equivalent to the evolution \eqref{highervfe} and \eqref{highervfe'}
on the curvature and torsion of $\pos(x)$. 

{\bf Remark:}
A different geometric derivation of the mKdV model \eqref{mkdvmodel} 
appears in work \cite{Anco06} on non-stretching flows of curves
in three-dimensional manifolds with constant curvature, i.e. 
${\rm S}^3$, ${\rm H}^3$, $\mathbb{R}^3$,
where the spin vector $S$ is identified with the components of 
the unit tangent vector in a moving frame defined by 
parallel transport along the curve. 
The mKdV model also has been derived in \cite{Fuchssteiner} 
as a higher-order symmetry of the Heisenberg model 
by non-geometric methods. 

All of these $SO(3)$ vector models
describe dynamical maps $\gamma$ on the unit sphere
${\rm S}^{2}\subset \mathbb{R}^{3}$
by means of the identifications:
\EQ
S_{t}\leftrightarrow \gamma_{t}, \quad
S_{x}\leftrightarrow \gamma_{x}, \quad
\partial_{x}+S(S_{x}\cdot)\leftrightarrow \nabla_{x}, \quad
S\wedge \leftrightarrow J=*
\label{Smap}
\endEQ
and thus
\EQs
&&
\nabla_{x}\gamma_{x}\leftrightarrow S_{xx}+|S_{x}|^{2}S, \quad
\nabla^{2}_{x}\gamma_{x}\leftrightarrow
S_{xxx}+|S_{x}|^{2}S_{x}+\frac{3}{2}(|S_{x}|^{2})_{x}S,
\label{Smap1}\\
&&
J\gamma_{x}\leftrightarrow S\wedge S_{x}, \quad
J\nabla_{x}\gamma_{x}\leftrightarrow S\wedge S_{xx} ,
\label{Smap2}\\
&&
g(\gamma_x,\gamma_x) = |\gamma_{x}|_g^{2} \leftrightarrow 
|S_x|^2 = S_x\cdot S_x
\endEQs
where $g$ denotes the Riemannian metric on the sphere ${\rm S}^2$
(given by restricting the Euclidean inner product in $\mathbb{R}^3$ 
to the tangent space of ${\rm S}^2 \subset \mathbb{R}^3$). 

In particular, 
the $SO(3)$ Heisenberg model yields 
the Schr\"odinger map equation \eqref{schrodingermap} on ${\rm S}^{2}$,
while the $SO(3)$ mKdV model \eqref{mkdvmodel} is identified with
\EQ
-\gamma_{t}=
\nabla^{2}_{x}\gamma_{x}+\frac{1}{2}|\gamma_{x}|_g^{2} \gamma_{x}
\label{mkdvmap}
\endEQ
which is a mKdV map equation on ${\rm S}^{2}$
(i.e. a dynamical map version of the potential mKdV equation). 

Thus, Theorem~\ref{uhierarchy} provides a geometric realization of
the hierarchies of integrable vector models and dynamical maps 
containing the Heisenberg model and the Schr\"odinger map
as well as their mKdV counterparts. 

\section{Geometric hierarchy of integrable vector models and dynamical maps}

In general, any non-stretching space curve flow \eqref{curveflow} 
can be written in terms of a Frenet frame \eqref{frenet}
by an \eom/ of the form 
\EQ
\pos_t=a T + b N +c B
\label{rflow}
\endEQ
such that 
\EQ
D_x a =\kappa b . 
\label{nostretch}
\endEQ
This relation between the tangential and normal components of the motion
arises due to the non-stretching property 
\EQ
|\pos_x|=1
\endEQ
by which the motion preserves the local arclength $ds=|\pos_x|dx$ 
of the space curve if (and only if)
$\pos_x\cdot\pos_{tx}=0$.
As a consequence, 
through the Serret-Frenet equations \eqref{xfrenet}--\eqref{K}, 
the tangent vector $T=\pos_x$ along the space curve
obeys the \eom/
\EQ
T_{t}=\hat f_1 N+\hat f_2 B =f \quad\perp\ T
\label{Tflow}
\endEQ
given by a linear combination of the normal and bi-normal vectors
with coefficients 
\EQ
\hat f_1 = D_x b -\tau c +\kappa a ,\quad
\hat f_2 = D_x c +\tau b .
\label{fcoeffs}
\endEQ

Now we consider a Hasimoto transformation 
\eqref{gaugetransform}--\eqref{parallelgauge}
from the Frenet frame \eqref{frenet} to a parallel frame \eqref{complexE}.
The \eom/ \eqref{rflow} on $\pos$ takes the form
\EQ
\pos_t
=h_\parallel E^\parallel +\Re( \bar h_\perp E^\perp)
= h_\parallel T +\Re(h_\perp e^{i\theta})N +\Im(h_\perp e^{i\theta})B
\label{rframeflow}
\endEQ
in terms of the tangential and normal components given by 
\EQ
h_\perp = (b+ic) e^{-i\theta} ,\quad
h_\parallel =a , 
\endEQ
with these components satisfying the relation \eqref{nostretch} given by 
\EQ
D_x h_\parallel = \Re(\bar u h_\perp)
\label{framenostretch}
\endEQ 
where $u=\kappa e^{-i\theta}$. 
Correspondingly, 
from the Frenet equations \eqref{complexEx} of the parallel frame, 
the \eom/ \eqref{Tflow} for the tangent vector $T=\pos_x$ has the form 
\EQ
T_{t} =\Re( \bar\varpi E^\perp)
= \Re( \varpi e^{i\theta} )N +\Im( \varpi e^{i\theta} )B
\label{Tframeflow}
\endEQ
in terms of 
\EQ
\varpi = D_x h_\perp + h_\parallel u
\label{weq}
\endEQ
which encodes the normal and bi-normal components
\EQ
\hat f_1 +i\hat f_2 =\varpi e^{i\theta} . 
\endEQ

The evolution of $T$ is thus specified by the variable $\varpi$, 
while the underlying evolution of $\pos$ is specified in terms of 
the variable $h_\perp$, with $h_\parallel$ given by the non-stretching 
condition \eqref{framenostretch}. 
From equation \eqref{weq} these variables are related by 
\EQ
\varpi = D_x h_\perp + u D_x^{-1}\Re(\bar u h_\perp) =\mathcal{J}(h_\perp) .
\label{Jeq}
\endEQ
The operator here 
\EQ
\mathcal{J} = D_x + u D_x^{-1}\Re(u \mathcal{C}) 
\label{Jop}
\endEQ
is related to the Hamiltonian operator $\mathcal{I}=-i$ by the properties
\EQ
-\mathcal{J} = \mathcal{R}^*\mathcal{I}^{-1} = \mathcal{I}^{-1}\mathcal{R} 
\quad\text{and}\quad
-\mathcal{J}^{-1} = \mathcal{R}^{-1}\mathcal{I} = \mathcal{I}\mathcal{R}^{*-1} , 
\endEQ
where $\mathcal{R}$ and $\mathcal{R}^*$ 
are the recursion operators \eqref{Rops}.
Consequently, 
$\mathcal{J}^{-1}$ is a formal Hamiltonian operator 
compatible with $\mathcal{I}$. 

\begin{proposition}\label{curveflowconditions}
The evolution \eqref{rflow} of a non-stretching space curve $\pos(x)$
can be expressed in terms of a geometrical variable that determines
the corresponding evolution \eqref{Tflow} of the tangent vector $T=\pos_x$
through the relation $T_t=(\pos_t)_x$. 
In particular, 
\EQ
h_\perp = \mathcal{J}^{-1}(\varpi) = \mathcal{R}^{-1}(i\varpi)
= i\mathcal{R}^{*-1}(\varpi)
\label{hperpeq}
\endEQ
yields the normal components of the evolution vector $\pos_t$ 
in a parallel frame,
where $\varpi$ represents the frame components of $T_t$. 
The curvature $\kappa$ and torsion $\tau$ of $\pos(x)$ correspondingly 
have the evolution
\EQs
\kappa_{t}&=& D_{x}{\hat f_1}-\tau{\hat f_2 }
=\Re( e^{i\theta}D_{x}\varpi)
\label{curvflow}\\
\tau_{t}&=&
D_{x}(\kappa^{-1}D_{x}{\hat f_2} +\tau \kappa^{-1}{\hat f_1})+\kappa{\hat f_2}
=D_{x}(\kappa^{-1}\Im(e^{i\theta}D_{x}\varpi)) +\kappa \Im(e^{i\theta}\varpi)
\label{torsflow}
\endEQs
which can be expressed in terms of the Frenet frame coefficients 
$a,b,c$ of $\pos_t$
through the relations \eqref{fcoeffs}. 
\end{proposition}

Conditions will now be stated 
within the general class of flows \eqref{flows} on $u$
such that the various evolutions 
\eqref{curvflow}--\eqref{torsflow}, \eqref{Tframeflow}, \eqref{rframeflow}, 
\eqref{Tflow},  \eqref{rflow}, and \eqref{uHamflow}--\eqref{Hamgrad}
each define a geometric flow. 

\begin{theorem}\label{geomflows}
For a non-stretching flow of a space curve $\pos(x)$ in $\mathbb{R}^3$,
the following conditions are equivalent:
\newline
(i) Its tangent vector $T=\pos_x=S$ obeys a $SO(3)$-invariant vector model 
iff $\hat f_1$ and $\hat f_2$ are functions of scalar invariants 
formed out of $S$ and its $x$ derivatives 
(modulo differential consequences of $S\cdot S=1$), i.e. 
\EQ
\hat f_1+i\hat f_2=
\hat{f}(S_{x}\cdot S_{x}, S_{x}\cdot S_{xx}, S_{xx}\cdot S_{xx}, \ldots, 
S_{xx}\cdot (S\wedge S_{x}), S_{xxx}\cdot (S\wedge S_{x}), 
S_{xxx}\cdot (S\wedge S_{xx}),\ldots) .
\label{}
\endEQ
(ii) Its principal normal component $u=T_{x}\cdot E^\perp$ 
in a parallel frame \eqref{complexE}
satisfies a $U(1)$-invariant evolution equation iff 
$\varpi$ is an equivariant function of $u$, $\bar{u}$, 
and $x$ derivatives of $u$ and $\bar{u}$,
under the action of a rigid ($x$-independent) $U(1)$ rotation group 
$u\rightarrow e^{-i\phi}u$ (with $\phi=$const.), i.e. 
\EQ
\varpi=u\hat{f}(|u|, |u|_x, |u|_{xx},\ldots,(\arg u)_x, (\arg u)_{xx},\ldots) .
\label{}
\endEQ
(iii) Its curvature $\kappa$ and torsion $\tau$ 
satisfy geometric evolution equations in terms of 
invariants and differential invariants of $\pos(x)$ iff 
$\varpi e^{i\theta}$ is a function of $\kappa$, $\tau$, 
and their $x$ derivatives, i.e.
\EQ
\varpi=
e^{-i\theta}\kappa
\hat{f}(\kappa,\kappa_x,\kappa_{xx},\ldots,\tau,\tau_x,\tau_{xx},\ldots) .
\label{}
\endEQ
(iv) Its \eom/ is invariant under the Euclidean isometry group 
$SO(3)\rtimes\mathbb{R}^3$ iff $a,b,c$ are scalar functions of
the curvature $\kappa$, torsion $\tau$, and their $x$-derivatives,
subject to the non-stretching condition \eqref{nostretch}. 
\end{theorem}

The proof of this proposition amounts to enumerating the Euclidean 
(differential) invariants of a space curve $\pos(x)$ 
with an arclength parameterization $x$, as shown in appendix~\ref{Proof}.

We are now able to derive the entire hierarchy of 
$SO(3)$-invariant vector models and geometric space curve motions
that correspond to all of the $U(1)$-invariant flows on $u$ 
in Theorem~\ref{uhierarchy}.

From equation \eqref{Tframeflow} combined with the hierarchy \eqref{Hamgrad},
the evolution of the spin vector $S=T=\pos_x$ can be written as
\EQ
S_t=\Re( \bar E^\perp \mathcal{R}^{*n}(u) ) ,\quad n=0,1,2,\ldots
\label{Sflow}
\endEQ
as generated via the recursion operator 
$\mathcal{R}^*=iD_x -u D_x^{-1}\Re(iu\mathcal{C})$. 
The main step is now to establish the operator identity
\EQ
\Re(\bar E^\perp \mathcal{R}^*) = \mathcal{S}\Re(E^\perp\mathcal{C})
\endEQ
where $\mathcal{S}$ is a spin vector operator 
corresponding to $\mathcal{R}^*$,
and $\Re(E^\perp\mathcal{C})$ is the operator that produces 
a vector $f$ in the perp space of $S$ in $\mathbb{R}^3$ when applied to 
the components of $f$ with respect to $E^\perp$ 
(i.e. $f=\Re(E^\perp\mathcal{C}\hat f)$ if $f\cdot S=0$, 
where $\hat f=f\cdot E^\perp$). 
Through the Frenet equations \eqref{complexEx} 
and the orthonormality relations \eqref{complexEnorms}
on $E^\perp$ and $E^\parallel$, 
we straightforwardly find 
\EQ
\Re(\bar E^\perp u)=E^\parallel_x = S_x
\endEQ
and 
\EQs
&&
\Re(\bar E^\perp iD_x\hat f)=
E^\parallel\wedge \Re(\bar E^\perp D_x\hat f)
=S\wedge D_x \Re(E^\perp\mathcal{C}\hat f) = S\wedge D_x f
\nonumber\\
&&
D_x^{-1}\Re(iu\mathcal{C}\hat f) = 
D_x^{-1}\Re((E^\parallel\wedge E^\parallel_x)\cdot E^\perp\mathcal{C}\hat f)
= D_x^{-1}\Re((S\wedge S_x) \cdot f)
\nonumber
\endEQs
for any vector $f(x)$, orthogonal to $S$ in $\mathbb{R}^3$,
with components ${\hat f}(x) =E^\perp\cdot f(x)$. 
Hence, this yields the vector operator 
\EQ
\mathcal{S} = S\wedge D_x - S_x D_x^{-1}(S\wedge S_x)\cdot
\label{Sop}
\endEQ

\begin{theorem}\label{Shierarchy}
The bi-Hamiltonian flows \eqref{ut} on $u(t,x)$ correspond to 
a hierarchy of integrable $SO(3)$-invariant vector models 
\EQ
S_t = \left( S\wedge D_x - S_x D_x^{-1}(S\wedge S_x)\cdot\ \right)^n S_x
=f^{(n)} ,
\quad n=0,1,2,\ldots\quad 
\label{modeleqs}
\endEQ
generated by the recursion operator \eqref{Sop}. 
These models have the equivalent geometrical formulation
\EQ
\gamma_t = 
\left( J\nabla_x - \gamma_x D_x^{-1} g(J\gamma_x,\ )\right)^n \gamma_x 
=F^{(n)} ,
\quad n=0,1,2,\ldots 
\label{mapeqs}
\endEQ
expressed as dynamical maps $\gamma(t,x)$ into the 2-sphere 
${\rm S}^2\subset\mathbb{R}^3$. 
Moreover, the Hamiltonians \eqref{Ham} for all the flows on $u(t,x)$
correspond to a set of constants of motion 
$\mathfrak{H}^{(0)}=\int_C H^{(0)} dx$, 
$\mathfrak{H}^{(1)}=\int_C H^{(1)} dx$, 
$\mathfrak{H}^{(2)}=\int_C H^{(2)} dx$, etc.
for each vector model \eqref{modeleqs}
and each dynamical map equation \eqref{mapeqs}. 
This entire set has the explicit form given by the densities 
(modulo total $x$-derivatives) 
\EQ\label{SHam}
(1+n) H^{(n)} = D_x^{-1}(S_x \cdot D_x f^{(n)})
=D_x^{-1} g(\gamma_x,\nabla_x F^{(n)}) ,
\quad n=0,1,2,\ldots 
\endEQ
which are scalar polynomials formed out of $SO(3)$-invariant 
wedge products and dot products of $S,S_x,S_{xx},\ldots$ 
in terms of the \esom/ for $S(t,x)$, 
or equivalently, 
scalar inner products of $\gamma_x,J\gamma_x,\nabla_x\gamma_x,J\nabla_x\gamma_x,\ldots$ 
formed in terms of the \esom/ for $\gamma(t,x)$.
\end{theorem}

The vector models \eqref{modeleqs} have been derived previously 
in \cite{BarFokPap}
by non-geometric methods (based on a Lax pair representation).
As one new result, 
Theorem~\ref{Shierarchy} 
derives the equivalent dynamical map equations \eqref{mapeqs},
along with their recursion operator, 
and provides an explicit expression for the constants of motion 
for all of these integrable generalizations of 
the Heisenberg spin model \eqref{heisenbergmodel}
and the mKdV spin model \eqref{mkdvmodel}. 
In particular, from the hierarchy \eqref{modeleqs}, 
higher-derivative versions of 
the Heisenberg spin model $n=1$ as given by $n=3,5,\ldots$ 
are seen to describe higher-derivative Schr\"odinger maps;
similarly, higher-derivative versions of 
the mKdV spin model $n=2$ as given by $n=4,6,\ldots$ 
are found to describe higher-derivative mKdV maps. 

{\bf Ex:} $n=3$ and $n=4$ respectively yield
a 4th order Heisenberg vector model
\EQ
S_t = S\wedge S_{xxxx} + \frac{5}{2}( |S_x|^2 S\wedge S_x )_x
\endEQ
and a 5th order mKdV vector model
\EQ
-S_t = 
S_{xxxxx} -\frac{5}{2}( |S_{xx}|^2 S + |S_x|^2 S_x 
-\frac{7}{4} |S_x|^4 S)_x + \frac{5}{2}( |S_x|^2 S )_{xxx}
\endEQ
which are described geometrically by 
a 4th order Schr\"odinger map equation
\EQ
-\gamma_t=
J\nabla_x^3 \gamma_x +\frac{1}{2}\nabla_x(|\gamma_x|_g^2 J\gamma_x)
-\frac{1}{2} g(J\gamma_x,\nabla_x\gamma_x)\gamma_x
\endEQ
and a 5th order mKdV map equation
\EQ
\gamma_t=
\nabla_x^4 \gamma_x +\frac{3}{4}(|\gamma_x|_g^2)_x \nabla_x\gamma_x
+( \frac{5}{4} (|\gamma_x|_g^2)_{xx} -\frac{3}{2} |\nabla_x\gamma_x|_g^2
+\frac{3}{8} |\gamma_x|_g^4 )\gamma_x .
\endEQ

{\bf Remark:}
The equations of motion of these dynamical maps $\gamma(t,x)$
{\em do not} locally preserve the arclength $ds=|\gamma_{x}|_g dx$
in the $x$ direction along $\gamma$ on ${\rm S}^2$, 
namely $|\gamma_x|_g$ has a nontrivial time evolution for any of 
the dynamical map equations \eqref{mapeqs}, 
and additionally, 
the total arclength $\int_C |\gamma_x|_g dx$ is time dependent. 

The spin vector recursion operator \eqref{Sop} has the factorization
\EQ\label{Sopfactorize}
\mathcal{S} =( D_x +S S_x\cdot\ +S_x D_x^{-1}S_x\cdot\ )( S\wedge\ )
\endEQ
where, as shown by the results in \cite{BarFokPap},
the operators 
$D_x +S S_x\cdot\ +S_x D_x^{-1}S_x\cdot$ and $(S\wedge\ )^{-1}=-S\wedge$
constitute a compatible Hamiltonian pair with respect to 
the spin vector variable $S(t,x)$. 
By comparison, using the present geometric framework, 
we now directly derive the explicit bi-Hamiltonian structure of 
the hierarchy of vector models \eqref{modeleqs}
through the bi-Hamiltonian flow equation \eqref{ut} on $u(t,x)$
in Theorem~\ref{uhierarchy}.

The starting point is the variational identity
\EQ\label{varH}
\delta H^{(n)} \equiv \delta S\cdot(\delta H^{(n)}/\delta S)
\equiv 2\Re(\bar\varpi^{(n)} \delta u)
\endEQ
holding modulo total $x$-derivatives 
and modulo (differential consequences of) $S\cdot\delta S=0$, 
with the Hamiltonian densities given in the equivalent forms 
\eqref{Ham} and \eqref{SHam},
and with $\varpi^{(n)}$ given by the hierarchy \eqref{nthHamgrad}
generated through the recursion operator $\mathcal{R}^*$.
To begin we derive an explicit expression for $\delta u$ 
in terms of $\delta S=\delta E^\parallel$
by taking the Frechet derivative of the Frenet equation \eqref{complexEx}
for the normal vectors in a parallel frame with respect to $u$:
\EQ\label{varEperpx}
D_x\delta E^\perp = -\delta u E^\parallel -u\delta E^\parallel . 
\endEQ
The $\bar E^\perp$ component of \eqref{varEperpx} gives
$$D_x(\bar E^\perp \cdot \delta E^\perp) 
= -u \bar E^\perp \cdot \delta E^\parallel - \bar u E^\parallel\cdot\delta E^\perp
= -2i\Re(i\bar u E^\perp \cdot \delta E^\parallel)$$
via $E^\parallel\cdot\delta E^\perp=-E^\perp \cdot \delta E^\parallel$
due to the orthogonality of $E^\perp,E^\parallel$. 
Similarly, the $E^\parallel$ component of \eqref{varEperpx} yields
$$\delta u = D_x(E^\perp \cdot \delta E^\parallel) 
+\frac{1}{2} u \bar E^\perp \cdot\delta E^\perp .$$
Combining these two expressions, we obtain 
\EQ
\delta u = 
D_x(E^\perp \cdot \delta E^\parallel) 
-iu D_x^{-1}\Re(i\bar u E^\perp \cdot \delta E^\parallel)
\endEQ
whence
\EQ\label{1ststep}
\Re(\bar\varpi^{(n)} \delta u) =
\Re(\bar\varpi^{(n)} D_x(E^\perp \cdot \delta E^\parallel))
+\Im(\bar\varpi^{(n)} u)D_x^{-1}\Re(i\bar u E^\perp \cdot \delta E^\parallel) .
\endEQ
Integration by parts on both terms in \eqref{1ststep} then yields
\EQs
\Re(\bar\varpi^{(n)} \delta u) 
&=&
-\Re(E^\perp \cdot \delta E^\parallel D_x \bar\varpi^{(n)})
-\Re(i\bar u E^\perp \cdot \delta E^\parallel D_x^{-1}\Im(\bar\varpi^{(n)} u))
\nonumber\\
&\equiv & 
- \delta E^\parallel\cdot\Re( \bar E^\perp \mathcal{H}(\varpi^{(n)}) )
\label{2ndstep}
\endEQs
modulo total $x$-derivatives. 
Next, after use of the relations 
$i\mathcal{R}^*=-\mathcal{H}$
and $i\bar E^\perp =E^\parallel\wedge \bar E^\perp$
which follow from \eqref{Rops} and \eqref{iE}, 
we get
\EQ\label{3rdstep}
\Re(\bar\varpi^{(n)} \delta u) \equiv
\delta E^\parallel\cdot( E^\parallel\wedge\Re(\bar E^\perp \varpi^{(n+1)}) ) . 
\endEQ
Hence, equating \eqref{3rdstep} and \eqref{varH} yields
$$ \delta S\cdot( \delta H^{(n)}/\delta S
- 2S\wedge \Re(\bar E^\perp \varpi^{(n+1)}) ) \equiv 0 $$
from which we obtain the variational relation
\EQ\label{Hamid}
-S\wedge(\delta H^{(n)}/\delta S) = 2\Re(\bar E^\perp \varpi^{(n+1)})
= 2\Re(\bar E^\perp \mathcal{R}^{*\,n+1}(u)) =2f^{(n+1)}
\endEQ
where the final equality comes from the \eom/ \eqref{Sflow} 
combined with the hierarchy \eqref{modeleqs}. 
This result \eqref{Hamid}, 
together with the factorization of the recursion operator \eqref{Sopfactorize},
leads to the following Hamiltonian structure. 

\begin{theorem}\label{SbiHam}
In terms of the Hamiltonian densities \eqref{SHam}, 
the hierarchy of vector models \eqref{modeleqs} for $S(t,x)$ 
has two Hamiltonian structures
\EQ
S_t = -S\wedge(\delta H^{(n-1)}/\delta S) = f^{(n)} ,\quad
n=1,2,\ldots
\label{1stSHam}
\endEQ
and, for $n\neq 1$, 
\EQ
S_t = D_x( \delta H^{(n-2)}/\delta S 
+S D_x^{-1}(S_x\cdot\delta H^{(n-2)}/\delta S) ) = f^{(n)} ,\quad
n=2,3,\ldots
\label{2ndSHam}
\endEQ
where
\EQ\label{SHamops}
S\wedge
\quad\text{and}\quad 
D_x +S S_x\cdot\ +S_x D_x^{-1}S_x\cdot
\endEQ
are a compatible pair of Hamiltonian operators. 
Correspondingly, 
the hierarchy of dynamical maps \eqref{mapeqs} on $\gamma(t,x)$
has the Hamiltonian structures
\EQs
\gamma_t &=& -J(\delta H^{(n-1)}/\delta\gamma) = F^{(n)} ,\quad
n=1,2,\ldots
\\
&=& \nabla_x(\delta H^{(n-2)}/\delta\gamma)
+\gamma_x D_x^{-1} g(\gamma_x,\delta H^{(n-2)}/\delta\gamma) ,\quad
n=2,3,\ldots
\endEQs
in terms of the compatible Hamiltonian operators 
\EQ\label{mapHamops}
J \quad\text{and}\quad 
\nabla_x +\gamma_x D_x^{-1}g(\gamma_x,\ )
\endEQ
as given by the geometrical identifications \eqref{Smap}. 
Moreover, these bi-Hamiltonian pairs \eqref{SHamops} and \eqref{mapHamops}
are geometrically equivalent to the pair of compatible symplectic 
(inverse Hamiltonian) operators 
$-\mathcal{I}^{-1}$ and $\mathcal{J}=-\mathcal{R}^*\mathcal{I}^{-1}$
with respect to the flow variable $u(t,x)$ in the hierarchy 
\eqref{ut}--\eqref{Ham} 
(cf. {\em Theorems~\ref{uhierarchy} and~\ref{Shierarchy}}).
\end{theorem}

These results provide an explicit geometrical formulation of
the abstract symplectic structure given in \cite{TerUhl}
for the Schr\"odinger map equation \eqref{schrodingermap}, i.e. ($n=1$)
\EQ\label{schrodinger1stHam}
\gamma_t = J\nabla_x\gamma_x 
= -J(\delta H^{(0)}/\delta\gamma) ,\quad
H^{(0)}= \frac{1}{2} g(\gamma_x,\gamma_x) ,
\endEQ
and its higher-order generalizations ($n=3,5,\ldots$); 
the Hamiltonian structure of the mKdV map, i.e. ($n=2$)
\EQs
\gamma_t &=& -\nabla^{2}_{x}\gamma_{x}-\frac{1}{2}|\gamma_{x}|_g^2 \gamma_{x}
= \nabla_x(\delta H^{(0)}/\delta\gamma)
+\gamma_x D_x^{-1} g(\gamma_x,\delta H^{(0)}/\delta\gamma) 
\\
&=& -J(\delta H^{(1)}/\delta\gamma) ,\quad
H^{(1)}= -\frac{1}{2} g(\nabla_x\gamma_x,J\gamma_x) ,
\endEQs
and its higher-order generalizations ($n=4,6,\ldots$)
has not appeared previously in the literature. 
Note the $n=1$ and $n=2$ vector models respectively yield the well-known
Hamiltonian structure of the Heisenberg spin model \eqref{heisenbergmodel},
\EQ\label{Heisenberg1stHam}
S_t = S\wedge S_{xx} = -S\wedge(\delta H^{(0)}/\delta S) ,\quad
H^{(0)} = \frac{1}{2} |S_x|^2 ,
\endEQ
and the mKdV spin model \eqref{mkdvmodel}, 
\EQs
S_t &=& -S_{xxx}-\frac{3}{2}(|S_x|^2 S)_x 
= D_x( \delta H^{(0)}/\delta S +S D_x^{-1}(S_x\cdot\delta H^{(0)}/\delta S) ) 
\\
&=& -S\wedge(\delta H^{(1)}/\delta S) ,\quad
H^{(1)} = \frac{1}{2} S_x\cdot(S\wedge S_{xx}) .
\endEQs

{\bf Remark:}
There is a second Hamiltonian structure for both 
the Schr\"odinger map equation and the Heisenberg spin model. 
This structure, in contrast to the first Hamiltonian structure
\eqref{schrodinger1stHam} and \eqref{Heisenberg1stHam}, 
turns out to involve a {\em non-polynomial} Hamiltonian density 
defined as follows. 
Let $\xi(\gamma)$ be a vector field on ${\rm S}^2$ such that
its divergence is constant at all points $\gamma\in {\rm S}^2$. 
(This property geometrically characterizes $\xi(\gamma)$ 
as a homothetic vector with respect to the metric-normalized volume form 
$\epsilon_g$ on ${\rm S}^2$, i.e.
$\mathcal{L}_\xi \epsilon_g= c\epsilon_g$ 
for some constant $c\neq 0$.)
Then the Hamiltonian density given by 
\EQ\label{2ndHam}
H^{(-1)} = g(\xi(\gamma),J\nabla\gamma_x) ,\quad
\div_g\xi(\gamma)=1
\endEQ
can be shown to satisfy (see Appendix~\ref{SchrodingerHamProof})
\EQ\label{2ndHamder}
\delta H^{(-1)} \equiv g(\delta\gamma,J\nabla\gamma_x)
\endEQ
modulo total $x$-derivatives, 
so thus
\EQ\label{Schrodinger2ndHam}
\gamma_t 
= \nabla_x(\delta H^{(-1)}/\delta\gamma)
+\gamma_x D_x^{-1} g(\gamma_x,\delta H^{(-1)}/\delta\gamma) 
= \nabla_x (J\gamma_x) 
\endEQ
yields a second Hamiltonian structure for 
the Schr\"odinger map \eqref{schrodingermap}. 
The corresponding second Hamiltonian structure for
the Heisenberg spin model \eqref{heisenbergmodel} is given by 
\EQ\label{Heisenberg2ndHam}
S_t = D_x( \delta H^{(-1)}/\delta S +S D_x^{-1}(S_x\cdot\delta H^{(-1)}/\delta S) ) 
= (S\wedge S_x)_x ,\quad
H^{(-1)} = \xi(S)\cdot(S\wedge S_x)
\endEQ
in terms of a vector function $\xi(S)$ such that 
\EQ
S \cdot \xi(S)=0, \quad
\partial^\perp_S\cdot \xi(S) =1 ,
\endEQ
where the operator 
$\partial^\perp_S = \partial_S -S (S\cdot\partial_S)$
is the orthogonal projection of 
a gradient with respect to the components of $S$. 
From the properties
\EQ\label{gradS}
S\cdot\partial^\perp_S =0 ,\quad
\partial^\perp_S (S\cdot S)=0 ,
\endEQ
the Hamiltonian density can be shown to satisfy 
\EQ\label{2ndHamS}
\delta H^{(-1)} \equiv \delta S\cdot(S\wedge S_x)
\endEQ
modulo total $x$-derivatives (see Appendix~\ref{HeisenbergHamProof}). 
Thus, both the Heisenberg spin model and the Schr\"odinger map equation 
are Hamiltonian \esom/ with respect to the corresponding bi-Hamiltonian pairs
\eqref{SHamops} and \eqref{mapHamops}. 

To conclude, 
a counterpart of Theorem~\ref{Shierarchy} will now be stated
for the underlying space curve motions on $\pos$. 

\begin{theorem}\label{curvehierarchy}
The $SO(3)$-invariant vector models \eqref{modeleqs} on $S(t,x)$
correspond to a hierarchy of integrable flows of non-stretching space curves
\EQ
\pos_t = a^{(n-1)}T +b^{(n-1)}N +c^{(n-1)}B ,\quad |\pos_x|=1, 
\quad n=1,2,\ldots\quad 
\label{curveeqs}
\endEQ
with geometric coefficients
\EQs
&
c^{(n)} = \Re(\mathcal{Q}^{n}\kappa) ,\quad
b^{(n)} = -\Im(\mathcal{Q}^{n}\kappa) ,
\label{normalcoeffs}\\
&
a^{(n)} = -D_x^{-1}(\kappa\Im(\mathcal{Q}^{n}\kappa)) 
\label{tangentialcoeff}
\endEQs
given in terms of the recursion operator 
\EQ
\mathcal{Q} = iD_x -\tau-\kappa D_x^{-1}(\kappa\Im)
=e^{i\theta}\mathcal{R}^*e^{-i\theta} ,
\label{Qop}
\endEQ
where $\kappa,\tau$ are the curvature and torsion of $\pos$.
The tangential coefficients \eqref{tangentialcoeff} yield a set of 
non-trivial constants of motion 
$\int_C a^{(1)} dx =-\int_C\frac{1}{2}\kappa^2 dx$, 
$\int_C a^{(2)} dx = \int_C \tau\kappa^2 dx$, etc.
for each curve flow \eqref{curveeqs} in the hierarchy
(under suitable boundary conditions), 
where $C=\mathbb{R}$ or $\rm S^1$ is the coordinate domain of $x$. 
\end{theorem}

The \esom/ \eqref{curveeqs} arise from 
writing the flow equation \eqref{rframeflow} on $\pos$
in terms of the hierarchy \eqref{Hamgrad} by means of 
the relation \eqref{hperpeq}.
This yields 
\EQs
&&
a^{(n)} = D_x^{-1}\Re(i\bar u \varpi^{(n-1)}) 
= - D_x^{-1}\Im(\bar u \varpi^{(n-1)}) ,
\\
&&
b^{(n)} = \Re(ie^{i\theta} \varpi^{(n-1)}) 
= -\Im(e^{i\theta} \varpi^{(n-1)}) , 
\\
&&
c^{(n)} = \Im(ie^{i\theta} \varpi^{(n-1)}) 
= \Re(e^{i\theta} \varpi^{(n-1)}) . 
\endEQs
The geometric form \eqref{normalcoeffs}--\eqref{tangentialcoeff}
for these coefficients is then obtained through the identity
\EQ
e^{i\theta} \varpi^{(n-1)} =e^{i\theta} \mathcal{R}^{*n-1} u
= (e^{i\theta} \mathcal{R}^* e^{-i\theta})^{n-1} e^{i\theta} u
\endEQ
combined with the Hasimoto transformation 
\EQ
\kappa=e^{i\theta} u ,\quad 
\tau = -\theta_x .
\endEQ

\section{Surfaces and spatial Hamiltonian curve flows}

The results in Theorem~\ref{uhierarchy} for 1+1 flows 
can be generalized in a natural way to 2+1 flows by considering 
surfaces $\pos(x,y)$ that are foliated by 
space curves with a parallel framing in $\mathbb{R}^3$.
Here $y$ will denote a coordinate assumed to be transverse to these curves,
and $x$ will denote the arclength coordinate along the curves, 
so thus 
\EQ\label{xnostretch}
|\pos_x|=1 .
\endEQ
In this setting, a parallel frame consists of a triple of unit vectors
whose derivatives along each $y={\rm const.}$ coordinate line 
on the surface $\pos(x,y)$
lie completely in either the tangent space or the normal plane of this line 
in $\mathbb{R}^3$. 
The explicit form of such a frame is given by the vectors
\EQs
&&
\widetilde{E}_1=T=\pos_x ,
\label{E1frame}\\
&&
\widetilde{E}_2 = \Re( e^{-i\theta}(N+iB) ) 
=|\pos_{xx}|^{-1}( \cos\theta\ \pos_{xx} +\sin\theta\ \pos_x\wedge \pos_{xx} ),
\label{E2frame}\\
&&
\widetilde{E}_3 = \Im( e^{-i\theta}(N+iB) ) 
=|\pos_{xx}|^{-1}( -\sin\theta\ \pos_{xx} +\cos\theta\ \pos_x\wedge \pos_{xx} ),
\label{E3frame}
\endEQs
with 
\EQ\label{Erot}
\theta_x = - |\pos_{xx}|^{-2} (\pos_x\wedge \pos_{xx})\cdot\pos_{xxx}
\endEQ
where $T,N,B$ respectively denote 
the unit tangent vector, unit normal and bi-normal vectors
of the $y={\rm const.}$ coordinate lines.  


In matrix notation
these frame vectors satisfy the Frenet equations
\EQ
\widetilde{\mathbf{E}}_x=\mathbf{U}\widetilde{\mathbf{E}},\quad
\widetilde{\mathbf{E}}_y=\mathbf{V}\widetilde{\mathbf{E}}
\label{xyfrenet}
\endEQ
given by 
\EQ
\widetilde{\mathbf{E}}=
\left(\begin{array}{ccc}
\widetilde{E}_1 \\ \widetilde{E}_2 \\ \widetilde{E}_3
\end{array}\right) ,\quad
\mathbf{U}
=\left(\begin{array}{ccc}
0 & u_2 & u_3\\ -u_2 & 0 & 0\\ -u_3 & 0 & 0
\end{array}\right) \in \mathfrak{so}(3),\quad
\mathbf{V}
=\left(\begin{array}{ccc}
0 & v_2 & v_3\\ -v_2 & 0 & v_1\\ -v_3 & -v_1 & 0
\end{array}\right) \in \mathfrak{so}(3),
\label{UV}
\endEQ
where 
\EQ\label{uv}
u_2+i u_3 =u 
= (\widetilde{E}_2 +i\widetilde{E}_3)\cdot \widetilde{E}_{1x} ,\quad
v_2+i v_3 =v 
= (\widetilde{E}_2 +i\widetilde{E}_3)\cdot \widetilde{E}_{1y} 
\endEQ
are complex scalar variables,
and 
\EQ
v_1=\frac{1}{2} (\widetilde{E}_3 +i \widetilde{E}_2 )\cdot
(\widetilde{E}_{2y}+i \widetilde{E}_{3y})
= \bar v_1 
\label{v1}
\endEQ
is a real scalar variable. 
Note, similarly to the case of space curves,
here we have 
\EQ
(\widetilde{E}_3 +i \widetilde{E}_2 )\cdot 
(\widetilde{E}_{2x}+i \widetilde{E}_{3x})
=0
\endEQ
which characterizes $\widetilde{\mathbf{E}}(x,y)$ as a parallel frame
adapted to the foliation of the surface $\pos(x,y)$
by $x$-coordinate lines, i.e.
$\widetilde{E}_{1x}\perp T$, 
$\widetilde{E}_{2x}\parallel T$, $\widetilde{E}_{3x}\parallel T$.
From \eqref{Erot} 
we see this choice of framing is geometrically unique
up to rigid rotations that act on the normal vectors \eqref{E2frame}--\eqref{E3frame}:
\EQ\label{framegroup}
\theta\rightarrow \theta +\phi ,\quad
\widetilde{E}_2+i \widetilde{E}_3 \rightarrow
\exp(-i\phi)(\widetilde{E}_2 +i \widetilde{E}_3) ,\quad
\text{with $\phi=$const.}\ .
\endEQ 
Under such rotations, 
$v_1$ is invariant, while $u$ and $v$ undergo a rigid phase rotation,
\EQ\label{U1uv}
u\rightarrow e^{-i\phi} u , \quad
v\rightarrow e^{-i\phi} v .
\endEQ

To proceed we need to write down the tangent vectors of $\pos(x,y)$
in terms of the parallel framing, 
\EQ\label{rxy}
\pos_x=\widetilde{E}_1 ,\quad
\pos_y= q_1 \widetilde{E}_1 + q_2 \widetilde{E}_2 + q_3 \widetilde{E}_3 . 
\endEQ
The respective projections of $\pos_y$ 
orthogonal and parallel to $\pos_x$ are given by the scalar variables
\EQ
q_2+i q_3 =q
= (\widetilde{E}_2 +i\widetilde{E}_3)\cdot \pos_y ,\quad
q_1=\widetilde{E}_1 \cdot \pos_y ,
\label{q}
\endEQ
where, under rigid rotations \eqref{framegroup}
on the normal vectors in the parallel frame,
$q_1$ is invariant while $q$ transforms by a rigid phase rotation,
\EQ\label{U1q}
q\rightarrow e^{-i\phi} q .
\endEQ

{\bf Remark:}
Due to the transformation properties \eqref{U1uv} and \eqref{U1q},
the variables $u$, $v$, $q$ represent $U(1)$-{\em covariants} of $\pos(x,y)$
as geometrically defined with respect to the $x$-coordinate lines,
where $U(1)$ is the equivalence group of rigid rotations \eqref{framegroup}
that preserves the form of the framing \eqref{E1frame}--\eqref{Erot}
for the surface $\pos(x,y)$.
These variables will be seen later 
(cf. Propositions~\ref{intrinsic} and~\ref{extrinsic}) 
to encode both the intrinsic and extrinsic surface geometry
of $\pos(x,y)$ in $\mathbb{R}^3$.

We will now show that all surfaces $\pos(x,y)$
with the $x$ coordinate satisfying 
the non-stretching property \eqref{xnostretch}
have a natural geometrical interpretation as 
a spatial Hamiltonian curve flow with respect the $y$ coordinate. 
This interpretation arises directly from the structure equations
satisfied by the parallel frame \eqref{E1frame}--\eqref{E3frame}
and the tangent vectors \eqref{rxy}
adapted to these coordinates. 
Firstly, 
the frame connection matrices given in \eqref{UV} obey
the zero-curvature condition
$\mathbf{U}_y-\mathbf{V}_x+[\mathbf{U},\mathbf{V}]=0$, 
which yields the structure equations
\EQ\label{1ststructeq}
u_y-v_x +iv_1 u =0 ,\quad
v_{1x}-\Im(\bar v u) =0 .
\endEQ
Secondly, 
the frame expansions of the tangent vectors given by \eqref{rxy}
obey the zero-torsion condition $(\pos_x)_y = (\pos_y)_x$,
leading to the structure equations
\EQ\label{2ndstructeq}
v-q_x -q_1 u =0 ,\quad
q_{1x}-\Re(\bar q u) =0 .
\endEQ

\begin{proposition}\label{spatialrflow}
Through the frame structure equations 
\eqref{1ststructeq} and \eqref{2ndstructeq},
the $U(1)$-covariants $u,v,q$ of $\pos(x,y)$ are related by 
the Hamiltonian equations
\EQ
u_y=\mathcal{H}(v) ,\quad
v=\mathcal{J}(q) , 
\endEQ
with 
$\mathcal{J}=-\mathcal{I}^{-1}\mathcal{R}=-\mathcal{R}^*\mathcal{I}^{-1}$,
where 
$\mathcal{H}=D_x -iu D_x^{-1}\Im(u\mathcal{C})$ and $\mathcal{I}=-i$
are a pair of compatible Hamiltonian operators with respect to $u$,
and 
$\mathcal{R}=\mathcal{H}\mathcal{I}^{-1}=i(D_x +u D_x^{-1}\Re(u\mathcal{C}))$
and 
$\mathcal{R}^*=\mathcal{I}^{-1}\mathcal{H}=iD_x -u D_x^{-1}\Re(iu\mathcal{C})$
are the associated recursion operators.
\end{proposition}

This leads to a main preliminary geometric result. 

\begin{lemma}\label{uvHamrel}
Let $u$ be an arbitrary complex-valued function of $x$ and $y$,
and define 
\EQs
&&
v=\mathcal{H}^{-1}(u_y) =i\mathcal{R}^{-1}(u_y) ,\quad
q=\mathcal{J}^{-1}(v)=-\mathcal{R}^{-2}(u_y) ,
\label{v}\\
&&
v_1=D_x^{-1}\Im(\bar v u) ,\quad 
q_1 =\Re(\bar q u)
\endEQs
in terms of the function $u$ 
and the formal inverse operators $\mathcal{H}^{-1}$ and $\mathcal{J}^{-1}$,
where $\mathcal{R}^{-1}$ is the formal inverse of 
the recursion operator $\mathcal{R} =\mathcal{H}i=i\mathcal{J}$. 
Then the matrix equations \eqref{xyfrenet} and the vector equations \eqref{rxy}
constitute a consistent linear system of 1st order PDEs that determine
a surface $\pos(x,y)$ with a parallel framing 
of the $y={\rm const.}$ coordinate lines
for which $x$ is the arclength. 
\end{lemma}

For a given function $u(x,y)$, 
the resulting surface $\pos(x,y)$ and frame $\widetilde{\mathbf{E}}(x,y)$
are unique up to Euclidean isometries.

\section{Surface flows and 2+1 soliton equations}

The derivation of bi-Hamiltonian soliton equations
from non-stretching space curve flows in Theorem~\ref{uhierarchy}
and their geometrical correspondence to 
integrable vector models and dynamical map equations 
in Theorems~\ref{Shierarchy} and~\ref{SbiHam}
will now be generalized to geometrically analogous surface flows 
in $\mathbb{R}^3$. 
This will mean we consider surfaces that are 
non-stretching along one coordinate direction 
yet stretching in all transverse directions.
(No assumptions will be place on the topology of the surface.)

Such surface flows $\pos(t,x,y)$ can be written naturally in terms of 
a parallel frame \eqref{E1frame}--\eqref{E3frame}
adapted to the non-stretching coordinate lines,
for which $x$ will be the arclength coordinate
and $y$ will be a transverse coordinate, 
by an \eom/ 
\EQ\label{tr}
\pos_t =  h_1 \widetilde{E}_1 + h_2 \widetilde{E}_2 + h_3 \widetilde{E}_3 ,
\quad |\pos_x|=1 ,
\endEQ
with 
\EQ\label{heq}
h_{1} =D_x^{-1}\Re(\bar u h) ,\quad 
h_2+i h_3 =h .
\endEQ
Here $h_1$ and $h$ respectively determine the components of the flow
that are tangential and orthogonal to the non-stretching $x$-coordinate lines
on the surface. 
The relation \eqref{heq} imposes the non-stretching property 
by which the flow \eqref{tr} preserves the local arclength $ds=|\pos_x|dx$
along these coordinate lines, 
with $u=u_2+i u_3$ given by the components of 
the parallel frame connection matrix with respect to $x$ 
from \eqref{UV}. 

The corresponding evolution of the parallel frame is given by 
the same matrix equations \eqref{Et}--\eqref{W} that govern 
a parallel framing of non-stretching flows of space curves.
Equivalently, in the complex variable notation \eqref{w}
for the components of the evolution matrix \eqref{W}, 
the evolution equations on the frame vectors consist of 
\EQ
\widetilde{E}_{1t}
=\Re( \varpi(\widetilde{E}_2-i\widetilde{E}_3) ) ,\quad
\widetilde{E}_{2t}+i \widetilde{E}_{3t}
= -\varpi\widetilde{E}_1 +\varpi_1 (\widetilde{E}_3 -i \widetilde{E}_2 ) ,
\label{tE123}
\endEQ
where
\EQ
\varpi=D_x h +\varpi_1 u ,\quad
\varpi_{1x} =\Re({\bar u} h) 
\endEQ
are given in terms of $h$ and $u$. 
These equations have the following Hamiltonian interpretation. 

\begin{proposition}\label{surfaceflow}
The respective evolution equations \eqref{tr} and \eqref{tE123} 
for the surface and the parallel frame 
are related by 
\EQ
h = \mathcal{J}^{-1}(\varpi) 
= \mathcal{R}^{-1}(i\varpi) = i\mathcal{R}^{*-1}(\varpi)
\label{hw}
\endEQ
where $\mathcal{J}^{-1}$ is a formal Hamiltonian operator
compatible with the Hamiltonian pair $\mathcal{H}$ and $\mathcal{I}$, 
and where
$\mathcal{R}^{-1}$, $\mathcal{R}^{*-1}$ 
are inverse recursion operators
(cf. {\em Proposition~\ref{spatialrflow}}).
\end{proposition}

Hence, 
surface flows of the type \eqref{tr}--\eqref{heq} 
can be expressed in terms of the complex scalar variable $\varpi$.
This variable also determines the resulting evolution of 
the frame connection matrices \eqref{UV} through the pair of 
zero-curvature equations
\EQs
&&
\mathbf{U}_t-\mathbf{W}_x+[\mathbf{U},\mathbf{W}]=0
\label{txcurv}\\
&&
\mathbf{V}_t-\mathbf{W}_y+[\mathbf{V},\mathbf{W}]=0
\label{tycurv}
\endEQs
which express the compatibility between the Frenet equations \eqref{xyfrenet}
and the evolution equations of the parallel frame 
in matrix form \eqref{Et}--\eqref{W}.
From \eqref{txcurv} 
we see $u$ satisfies the same evolution equation \eqref{uflow}
in terms of $\varpi$ as holds for non-stretching space curve flows
in Proposition~\ref{uHamstructure}. 
We then find that the evolution equation obtained from \eqref{tycurv}
holds identically as a consequence of Lemma~\ref{uvHamrel}. 
This result leads to the following Hamiltonian structure
derived from surface flows \eqref{tr}--\eqref{heq}.

\begin{lemma}\label{2DuHamstruct}
The Hamiltonian structure \eqref{uHamflow}--\eqref{Hamfunct} for 1+1 flows 
generalizes to 2+1 flows on $u(t,x,y)$ 
given by Hamiltonians of the form 
\EQ
\mathfrak{H}=\iint_C 
H(x,y,u,\bar u,u_x,\bar u_x,u_y,\bar u_y,u_{xx},\bar u_{xx},
u_{xy},\bar u_{xy},u_{yy}, \bar u_{yy},\ldots) dxdy 
\endEQ
(where $C$ denotes the coordinate domain of $(x,y)$).
In particular, 
\EQ
\varpi=\delta H/\delta \bar u
\label{}
\endEQ
yields a 2+1 Hamiltonian evolution equation 
\EQ
u_t =\mathcal{H}(\varpi) ,\quad
\mathcal{H}=D_x -iu D_x^{-1}\Im(u\mathcal{C}) ,
\endEQ
corresponding to a surface flow given by \eqref{tr}, \eqref{heq}, \eqref{hw}. 
\end{lemma}

Since the Hamiltonian operator $\mathcal H$
does not contain the $y$ coordinate,
it obviously has $y$-translation symmetry.
Hence starting from $-u_y\partial/\partial u$,
there will be a hierarchy of commuting Hamiltonian vector fields
\EQ
i\varpi^{(n)}\partial/\partial u =-\mathcal{R}^n (u_y)\partial/\partial u ,
\quad n=0,1,2,\ldots
\label{2DRflow}
\endEQ
where
\EQ
\varpi^{(n)}=\delta H^{(n)}/\delta \bar u=\mathcal{R}^{*n}(iu_y) ,
\quad n=0,1,2,\ldots
\label{2DRHam}
\endEQ
are derivatives of Hamiltonian densities $H^{(n)}$. 
An explicit expression for these densities can be derived by applying
the scaling symmetry methods in \cite{Anco03,Anco08}.

\begin{theorem}\label{2Duhierarchy}
The recursion operator 
$\mathcal{R}=\mathcal{H}\mathcal{I}^{-1}=i(D_x +u D_x^{-1}\Re(u\mathcal{C}))$
produces a hierarchy of integrable bi-Hamiltonian 2+1 flows
\EQ
u_t=\mathcal{H}(\delta \mathfrak{H}^{(n-1)}/\bar u)
=\mathcal{I} (\delta \mathfrak{H}^{(n)}/\delta \bar u),
\quad n=1,2,\ldots
\label{uHam2Dflow}
\endEQ
(called the {\em $+n$ flow})
in terms of the compatible Hamiltonian operators
$\mathcal{H}$ and $\mathcal{I}$,
with the Hamiltonians 
$\mathfrak{H}^{(n)}=\iint_C H^{(n)} dxdy$
given by
\EQ
H^{(n)}=\frac{-2}{1+n}
D_x^{-1}\Re({\bar u} \mathcal{R}^{n+1}(u_y))
,\quad n=0,1,2,\ldots
\label{Ham2Dflow}
\endEQ
(modulo total $x,y$-derivatives).
\end{theorem}

At the bottom of this hierarchy,
\EQ
H^{(0)}=\Re( i{\bar u}u_y) ,\quad
\delta H^{(0)}/\delta\bar u = iu_y = \varpi^{(0)}
\endEQ
yields the $+1$ flow
\EQ
-iu_t=u_{xy}+\nu_0 u ,\quad
\nu_{0x} = |u| |u|_y . 
\endEQ
This is a 2+1 nonlocal bi-Hamiltonian NLS equation 
which was first derived from Lax pair methods 
by Zhakarov \cite{Zha}
and Strachan \cite{Str92}.

Next in the hierarchy is
\EQ
H^{(1)}=\Re( {\bar u}_x u_y ) -\frac{1}{2}\nu_0 |u|^2 ,\quad
\delta H^{(1)}/\delta \bar u = -( u_{xy}+\nu_0 u ) = \varpi^{(1)} . 
\endEQ
This yields the +2 flow
\EQ
-u_t=u_{xxy}+|u|^2 u_y +\nu_0 u_x +i\nu_1 u ,\quad
\nu_{1x} = \Im(\bar u_y u_x)
\endEQ
which is a 2+1 nonlocal bi-Hamiltonian mKdV equation.
It can be written in the equivalent form
\EQ
-u_t=u_{xxy}+(\nu_0 u)_x +i\nu_2 u ,\quad
\nu_{2x} = \Im(\bar u u_{xy})
\endEQ
studied in work of Calogero \cite{Cal} and Strachan \cite{Str93}. 

These 2+1 flow equations have the following integrability properties. 

\begin{proposition}\label{2Dintegrable}
The hierarchy \eqref{uHam2Dflow}--\eqref{Ham2Dflow} displays
$U(1)$-invariance under phase rotations
$u\rightarrow e^{i\lambda}u$
and homogeneity under scalings
$x\rightarrow\lambda x$, $y\rightarrow\lambda y$,
$u\rightarrow \lambda^{-1}u$, 
with $t\rightarrow \lambda^{2+n}t$ for the $+n$ flow, 
where the scaling weight of $H^{(n)}$ is $-3-n$.
Each of the evolution equations \eqref{uHam2Dflow} in this hierarchy
admits the constants of motion
\EQ
D_t\iint_C i\bar u u_y\, dxdy 
=0 ,\quad
D_t\iint_C \bar u_x u_y -\frac{1}{2} \nu_0 |u|^2 dxdy
=0 ,\quad\ldots
\endEQ
and
\EQ\label{2Dcom}
D_t\iint_C |u|^2 dxdy 
=0 ,\quad
D_t\iint_C i{\bar u} u_x\, dxdy 
=0 ,\quad
D_t\iint_C |u_x|^2 -\frac{1}{4} |u|^4 dxdy 
=0 ,\quad\ldots
\endEQ
comprising, respectively, 
all of the 2+1 Hamiltonians \eqref{Ham2Dflow} 
plus the 1+1 Hamiltonians \eqref{Ham} extended to two spatial dimensions
(under suitable boundary conditions depending on the coordinate domain 
$C$ of $(x,y)$).
Additionally, 
these evolution equations \eqref{uHam2Dflow} each admit 
the corresponding Hamiltonian symmetries  
\EQ
-u_y\partial/\partial u, \quad
-i(u_{xy}+\nu_0 u)\partial/\partial u ,\quad
\ldots
\endEQ
plus
\EQ\label{2Dsymm}
i u\partial/\partial u, \quad
-u_x\partial/\partial u, \quad
-i(u_{xx}+\frac{1}{2}|u|^2u)\partial/\partial u ,\quad
\ldots\ .
\endEQ
\end{proposition}

We note the constants of motion \eqref{2Dcom} and symmetries \eqref{2Dsymm}
are inherited from the 1+1 integrability properties in Theorem~\ref{uhierarchy}
as a consequence of the fact that the Hamiltonian phase-rotation 
vector field $iu\partial/\partial u$ 
(which generates the hierarchy of 1+1 flows \eqref{ut})
commutes with the Hamiltonian $y$-translation 
vector field $-u_y\partial/\partial u$
(which generates the hierarchy of 2+1 flows \eqref{uHam2Dflow}). 

\section{2+1 vector models and dynamical maps}

Each evolution equation \eqref{uHam2Dflow} in the hierarchy 
presented in Theorem~\ref{2Duhierarchy}
determines a surface flow $\pos(t,x,y)$ 
and a corresponding 2+1 vector model for $S(t,x,y)= \pos_x=\widetilde{E}_1$
through the frame evolution equations \eqref{tE123}
in a similar manner to the derivation of flows of space curves
and 1+1 vector models.

The $+1$ flow yields the geometric $SO(3)$ vector model
known as the M-I equation \cite{Myr,Myr1,Myr4}
\EQ
S_t = S\wedge S_{xy} +v_1 S_x = (S\wedge S_y +v_1 S)_x ,\quad
v_{1x} = -S\cdot (S_x\wedge S_y) ,
\label{2DHeisenbergmodel}
\endEQ
which is a 2+1 integrable generalization of the $SO(3)$ Heisenberg model.
It corresponds to the surface flow
\EQ
\pos_t = \pos_x\wedge \pos_{xy} +v_1 \pos_x ,\quad
v_{1x} = -\pos_x\cdot(\pos_{xx} \wedge \pos_{xy}) ,\quad 
|\pos_x|=1 .
\endEQ
This flow equation describes the motion of 
a sheet of non-stretching filaments in Euclidean space, 
in analogy with the form of the vortex filament equations 
\eqref{filamentmotion}. 
Some properties of the model \eqref{2DHeisenbergmodel} have been studied 
recently in \cite{ZhaDenZhaWu,CheZho}. 

The $+2$ flow produces a 2+1 integrable generalization of 
the geometric $SO(3)$ mKdV model, 
\EQ
-S_t=S_{xxy}+ ( (S_x\cdot S_y +v_0)S -v_1 S\wedge S_x )_x ,\quad
v_{0x} = |S_x| |S_x|_y ,
\label{2Dmkdvmodel}
\endEQ
(the so-called M-XXIX equation \cite{Myr4}) 
which describes the surface flow
\EQ
-\pos_t
= \pos_{xxy} + (\pos_{xx}\cdot \pos_{xy} +v_0) \pos_x
-v_1 \pos_x\wedge \pos_{xx} ,\quad
v_{0x} = |\pos_{xx}| |\pos_{xx}|_y ,\quad |\pos_x|=1 . 
\endEQ

Each of these surface flows $\pos(t,x,y)$ in $\mathbb{R}^3$ 
geometrically corresponds to a dynamical map
$\gamma(t,x,y)$ on the unit sphere ${\rm S}^2 \subset\mathbb{R}^3$ 
through extending the identifications 
\eqref{Smap1} and \eqref{Smap2} as follows:
\EQ
S_{y}\leftrightarrow \gamma_{y} ,\quad
\partial_{y}+S(S_{y}\cdot)\leftrightarrow \nabla_{y} , 
\label{Sy}
\endEQ
\EQ
S_{xy}+(S_{x}\cdot S_{y})S
\leftrightarrow \nabla_{x}\gamma_{y} =\nabla_{y}\gamma_{x} ,
\label{Sxy}
\endEQ
\EQ
S\wedge S_{xy} \leftrightarrow J\nabla_{x}\gamma_{y} = J\nabla_{y}\gamma_{x} .
\label{SSxy}
\endEQ
The $+1$ and $+2$ flows thereby yield, respectively,
\EQ
\gamma_{t}=
J\nabla_{x} \gamma_{y} +v_1 \gamma_{x} ,\quad
v_{1x} = g(\gamma_{x},J\gamma_{y}) ,
\label{2DSchrodingermap}
\endEQ
and
\EQ
-\gamma_{t}=
\nabla^{2}_{x}\gamma_{y}+|\gamma_{x}|_g^2 \gamma_{y}
-v_1 J\nabla_{x} \gamma_{x} -v_2 \gamma_{x} ,\quad
v_{2x} = g(\gamma_{y},\nabla_{x} \gamma_{x}) ,
\label{2Dmkdvmap}
\endEQ
which are new nonlocal 2+1 integrable generalizations of 
the Schr\"odinger map equation \eqref{schrodingermap} on ${\rm S}^{2}$ 
and the mKdV map equation \eqref{mkdvmap} on ${\rm S}^{2}$.

The complete hierarchy of vector models and dynamical map equations 
in 2+1 dimensions
can be written down in the same manner as in 1+1 dimensions 
(cf. Theorems~\ref{Shierarchy} and~\ref{SbiHam})
by means of the spin vector recursion operator \eqref{Sop}
and its Hamiltonian factorization \eqref{Sopfactorize}. 
In particular, the obvious $y$-translation invariance of this operator
provides the geometric origin for the 2+1 generalization of 
the Heisenberg spin model and the mKdV spin model. 

\begin{theorem}\label{2DShierarchy}
(i) 
The bi-Hamiltonian flows \eqref{uHam2Dflow} correspond to 
a hierarchy of integrable $SO(3)$-invariant vector models in 2+1 dimensions
\EQ
S_t = \left( S\wedge D_x - S_x D_x^{-1}(S\wedge S_x)\cdot\ \right)^n S_y
=f^{(n)} ,
\quad n=1,2,\ldots 
\label{2Dmodeleqs}
\endEQ
which are geometrically equivalent to 2+1 dimensional 
dynamical maps $\gamma$ into the 2-sphere ${\rm S}^2\subset\mathbb{R}^3$
\EQ
\gamma_t = 
\left( J\nabla_x - \gamma_x D_x^{-1} g(J\gamma_x,\ )\right)^n \gamma_y
=F^{(n)} ,
\quad n=1,2,\ldots\ .
\label{2Dmapeqs}
\endEQ
(ii) 
Each vector model and dynamical map 
in the hierarchy \eqref{2Dmodeleqs}--\eqref{2Dmapeqs}
possesses a set of polynomial constants of motion that correspond to 
all of the Hamiltonians \eqref{Ham2Dflow} 
for the $+1,+2,\ldots$ flows \eqref{uHam2Dflow}, i.e.
$\mathfrak{H}^{(0)}=\iint_C H^{(0)} dxdy$, 
$\mathfrak{H}^{(1)}=\iint_C H^{(1)} dxdy$, 
etc., 
as obtained from the Hamiltonian densities 
(modulo total $x,y$-derivatives)
\EQ
(1+n) H^{(n)} = 
- D_x^{-1}( S_x\cdot D_x f^{(n+1)} ) 
= - D_x^{-1} g(\gamma_x,\nabla_x F^{(n+1)}) ,
\quad n=0,1,2,\ldots
\label{2DHam}
\endEQ
given in terms of the \esom/ \eqref{2Dmodeleqs} for $S(t,x)$ 
and \eqref{2Dmapeqs} for $\gamma(t,x)$. 
In addition, 
the vector models \eqref{2Dmodeleqs} and dynamical maps \eqref{2Dmapeqs}
each possess two non-polynomial constants of motion
$\mathfrak{H}^{(-1)}=\iint_C H^{(-1)} dxdy$ 
and 
$\mathfrak{H}^{(-2)}=\iint_C H^{(-2)} dxdy$
explicitly given by 
\EQs
&&
H^{(-2)} = \frac{1}{2} \xi(S)\cdot(S\wedge S_y)
=\frac{1}{2} g(\xi(\gamma),J\gamma_y) ,
\label{2D1stHam}\\
&&
H^{(-1)} = \frac{1}{2} S_x\cdot S_y+ \frac{1}{2} v_1 \xi(S)\cdot(S\wedge S_x)
= \frac{1}{2} g(\gamma_x,\gamma_y)+\frac{1}{2} v_1 g(\xi(\gamma),J\gamma_x) ,
\label{2D2ndHam}
\endEQs
where $\xi(\gamma)$ is a vector field with covariantly-constant divergence 
$\div_g\xi(\gamma)=1$
at all points $\gamma\in {\rm S}^2$, 
and where $\xi(S)$ is an analogous vector function satisfying 
$S \cdot \xi(S)=0$, $\partial^\perp_S\cdot \xi(S) =1$, 
in terms of the component-wise gradient operator 
$\partial^\perp_S = \partial_S -S (S\cdot\partial_S)$
with properties \eqref{gradS}. 
These Hamiltonian densities \eqref{2D1stHam}--\eqref{2D2ndHam}
correspond to two compatible nonlocal Hamiltonian structures for the $+0$ flow
\EQ
u_t=\mathcal{E}(\delta \mathfrak{H}^{(-2)}/\bar u)
=\mathcal{H} (\delta \mathfrak{H}^{(-1)}/\delta \bar u)
=-u_y 
\endEQ
with (cf. {\em Lemma~\ref{uvHamrel}})
\EQ
-\delta H^{(-1)}/\bar u=\mathcal{R}^{*-1}(iu_y)=v ,\quad
-\delta H^{(-2)}/\bar u=\mathcal{R}^{*-2}(iu_y)=-iq .
\endEQ
(iii)
In terms of the Hamiltonian densities 
\eqref{2DHam}, \eqref{2D1stHam} and \eqref{2D2ndHam}, 
all the 2+1 vector models \eqref{2Dmodeleqs} 
and dynamical map equations \eqref{2Dmapeqs} 
have the bi-Hamiltonian structure
\EQs\label{SHamstruct}
S_t &=& -S\wedge(\delta H^{(n-2)}/\delta S) 
= D_x( \delta H^{(n-3)}/\delta S 
+S D_x^{-1}(S_x\cdot\delta H^{(n-3)}/\delta S) ) 
\\ &=& f^{(n)} ,\quad 
n=1,2,\ldots
\nonumber
\endEQs
and 
\EQs\label{mapHamstruct}
\gamma_t &=& -J(\delta H^{(n-2)}/\delta\gamma) 
= \nabla_x(\delta H^{(n-3)}/\delta\gamma)
+\gamma_x D_x^{-1} g(\gamma_x,\delta H^{(n-3)}/\delta\gamma) 
\\ &=& F^{(n)} ,\quad 
n=1,2,\ldots
\nonumber
\endEQs
given by the respective pairs \eqref{SHamops} and \eqref{mapHamops}
of compatible Hamiltonian operators. 
\end{theorem}

{\bf Remark:}
Explicit bi-Hamiltonian formulations for 
the 2+1 generalization of the Heisenberg spin model \eqref{2DHeisenbergmodel}
and the geometrically corresponding 
new 2+1 integrable Schr\"odinger map \eqref{2DSchrodingermap}
are given by 
\EQ\label{2DHeisenbergHam}
S_t = -S\wedge(-S_{xy}+v_1 S\wedge S_x)
= D_x( S\wedge S_y +S D_x^{-1}(S_x\cdot (S\wedge S_y)) )
\endEQ
and 
\EQ\label{2DSchrodingerHam}
\gamma_t = -J(-\nabla_y\gamma_x +v_1 J\gamma_x)
= \nabla_x(J\gamma_y) +\gamma_x D_x^{-1} g(\gamma_x,J\gamma_y)
\endEQ
where
\EQs
&&
\delta H^{(-2)}=\delta S\cdot(S\wedge S_y) =g(\delta\gamma,J\gamma_y) 
\label{2D1stHamder}\\
&&
\delta H^{(-1)}=\delta S\cdot(-S_{xy}+v_1 S\wedge S_x)
=g(\delta\gamma,-\nabla_y\gamma_x +v_1 J\gamma_x) 
\label{2D2ndHamder}
\endEQs
yield the respective derivatives of the non-polynomial Hamiltonian densities 
\eqref{2D1stHam} and \eqref{2D2ndHam}.
These two densities in addition to all the polynomial densities \eqref{2DHam}
give a set of constants of motion for the Hamiltonian equations
\eqref{2DHeisenbergHam} and \eqref{2DSchrodingerHam}.
In particular, the first four constants of motion are explicitly given by 
the integrals
\EQs
&&
D_t \iint_C  \xi(S)\cdot(S\wedge S_y) dxdy
=D_t \iint_C  g(\xi(\gamma),J\gamma_y) dxdy
=0 ,
\label{nonpolycom}\\
&&
D_t \iint_C  S_x\cdot S_y + v_1 \xi(S)\cdot(S\wedge S_x) dxdy
\nonumber\\
&&
= D_t \iint_C  g(\gamma_x,\gamma_y)+ v_1 g(\xi(\gamma),J\gamma_x) dxdy
=0, 
\\
&&
D_t \iint_C -S\cdot(S_x \wedge S_{xy}) +v_1 |S_x|^2 dxdy
\nonumber\\
&&
= D_t \iint_C g(\nabla_y\gamma_x,J\gamma_x)+v_1 |\gamma_x|_g^2 dxdy 
=0 ,
\\
&&
D_t \iint_C S_{xx}\cdot S_{xy}- |S_x|^2(S_x\cdot S_y+\frac{1}{2} v_0)
-v_1 S\cdot(S_x \wedge S_{xx}) dxdy
\nonumber\\
&&
= D_t \iint_C  g(\nabla_y\gamma_x,\nabla_x\gamma_x)
-\frac{1}{2}v_0 |\gamma_x|_g^2 -v_1 g(J\gamma_x,\nabla_x\gamma_x) dxdy 
=0 
\endEQs
under suitable boundary conditions
(where $C$ denotes the coordinate domain of $(x,y)$).
The vector field $\xi(\gamma)$ on $\rm S^2$, 
or equivalently the vector function $\xi(S)$, 
in the non-polynomial constant of motion \eqref{nonpolycom}
has the geometrical meaning of a homothetic vector 
with respect to the metric-normalized volume form 
$\epsilon_g$ on ${\rm S}^2$, i.e.
$\mathcal{L}_\xi \epsilon_g= \epsilon_g$. 

Theorem~\ref{2DShierarchy} is established as follows. 
In parts~(i) and~(ii), 
the derivation of the \esom/ \eqref{2Dmodeleqs} and Hamiltonians \eqref{2DHam}
for $S(t,x)$ involves combining the evolution equation \eqref{tE123}
for the frame vector $\widetilde E_1 = S$
with the hierarchy \eqref{2DRflow} for the variable $\varpi$
by means of the identities
\EQs
&&
u (\widetilde{E}_2-i\widetilde{E}_3) = S_x -i S\wedge S_x ,
\\
&&
u_y (\widetilde{E}_2-i\widetilde{E}_3) =
-( S\wedge\mathcal{S}(S_y) + i\mathcal{S}(S_y) ) ,
\endEQs
in addition to 
\EQs
&&
\Re(\bar u \hat f) = S_x\cdot f ,\quad
\Im(\bar u \hat f) = (S\wedge S_x)\cdot f ,
\\
&&
\Re(\bar u_y \hat f) = \mathcal{S}(S_y)\cdot(S\wedge f) ,\quad
\Im(\bar u_y \hat f) = \mathcal{S}(S_y)\cdot f
\endEQs
holding for vectors $f$ in $\mathbb{R}^3$,
with components $\hat f=(\widetilde{E}_2+i\widetilde{E}_3)\cdot f$,
such that $f\cdot S=0$; 
here $\mathcal{S}$ is the recursion operator \eqref{Sop}.
Similarly,
the derivation of the Hamiltonian structures \eqref{SHamstruct} 
in part~(iii) for $n\neq 1$ 
relies on applying the previous identities to 
the bi-Hamiltonian structure \eqref{uHam2Dflow} 
for the flow equations on $u(t,x)$.  
The $n=1$ case reduces to computing the Hamiltonian derivatives 
\eqref{2D1stHamder}--\eqref{2D2ndHamder},
which is carried out in 
appendices~\ref{SchrodingerHamProof} and~\ref{HeisenbergHamProof}.
Finally, all of the corresponding results for $\gamma(t,x)$ 
are an immediate consequence of the geometric identifications
\eqref{Sy}--\eqref{SSxy} in addition to \eqref{Smap1}--\eqref{Smap2}.

\section{Geometric formulation}

There is a natural geometric formulation for the surface flows \eqref{tr}
corresponding to the 2+1 vector models \eqref{2Dmodeleqs}
and 2+1 dynamical maps \eqref{2Dmapeqs} in Theorem~\ref{2DShierarchy}.
We begin by writing down the the intrinsic and extrinsic surface geometry
of $\pos(x,y)$ in terms of the variables $u,v,q,v_1,q_1$ 
appearing in the structure equations of the parallel framing for
the non-stretching $x$-coordinate lines. 

\begin{proposition}\label{intrinsic}
Let $\pos(x,y)$ in $\mathbb{R}^3$ be a surface with a parallel framing
\eqref{xnostretch}--\eqref{Erot} adapted to the $x$ coordinate lines,
satisfying the structure equations \eqref{1ststructeq} and \eqref{2ndstructeq}.
Then, on the surface $\pos(x,y)$,
the infinitesimal arclength is given by the line element
\EQ\label{metric}
ds^2 = (dx+ q_1 dy)^2 + |q|^2 dy^2 ,
\endEQ
and the infinitesimal surface area is given by the area element
\EQ
dA = |q| dx\wedge dy .
\endEQ
\end{proposition}

All other aspects of the intrinsic surface geometry can be derived from
the line element \eqref{metric}. In particular, 
the 1st fundamental form (i.e. the surface metric tensor) is simply
$dx dx +2 q_1 dx dy + |q|^2 dy dy$,
from which the Gauss curvature can be directly computed in terms of the
$x,y$ coordinates \cite{Gug}. 

The extrinsic surface geometry can be determined through 
the surface normal vector
\EQ\label{surfnormal}
\vec{n} = \pos_x\wedge\pos_y 
=\Im(\bar q (\widetilde{E}_2 +i\widetilde{E}_3)) ,\quad |\vec{n}|=|q| ,
\endEQ
as given by the expression \eqref{rxy} for the surface tangent vectors
$\pos_x$ and $\pos_y$ in terms of the parallel frame 
along the $x$-coordinate lines.
This normal vector \eqref{surfnormal} depends on a choice of the 
transverse coordinate $y$ due to its normalization factor $|q|$.
To proceed, 
we use the following natural geometric framing \cite{Gug}
that is defined entirely by the non-stretching direction on the surface 
and the orthogonal direction of the surface unit-normal in $\mathbb{R}^3$:
\EQs
&&
e^\parallel = \pos_x 
= \widetilde{E}_1 ,
\label{ex}\\
&&
e^\perp =|\vec{n}|^{-1} \vec{n}  
= \Im(e^{i\psi} (\widetilde{E}_2 +i\widetilde{E}_3)) ,
\label{en}\\
&&
*e^\parallel =|\vec{n}|^{-1} \vec{n} \wedge \pos_x
= \Re(e^{i\psi} (\widetilde{E}_2 +i\widetilde{E}_3)) ,
\label{e*}
\endEQs
where 
\EQ\label{erot}
\psi=\arg(q) .
\endEQ
(Here $*$ denotes the Hodge dual acting in the tangent plane at each point
on the surface.)
Note this frame \eqref{ex}--\eqref{e*} differs from a parallel frame 
by a rotation through the angle \eqref{erot} 
applied to the frame vectors in the normal plane 
relative to the $x$-coordinate lines, 
with $e^\parallel$ and $*e^\parallel$ being an orthogonal pair of 
unit-tangent vectors on the surface $\pos(x,y)$,
and $e^\perp$ being a unit-normal vector for the surface. 

The Frenet equations of the frame $e^\parallel,*e^\parallel,e^\perp$
directly encode the extrinsic geometry of the surface $\pos(x,y)$.
In matrix notation, the Frenet equations with respect to the $x,y$ coordinates
are given by 
\EQ
\left(\begin{array}{ccc}
e^\parallel \\ *e^\parallel \\  e^\perp 
\end{array}\right)_x
=\left(\begin{array}{ccc}
0 & \alpha & \beta \\ -\alpha & 0 & \gamma\\ -\beta & -\gamma & 0
\end{array}\right) 
\left(\begin{array}{ccc}
e^\parallel \\ *e^\parallel \\ e^\perp 
\end{array}\right) ,\quad
\left(\begin{array}{ccc}
e^\parallel \\ *e^\parallel \\  e^\perp 
\end{array}\right)_y
=\left(\begin{array}{ccc}
0 & \mu & \rho \\ -\mu & 0 & \sigma\\ -\rho & -\sigma & 0
\end{array}\right) 
\left(\begin{array}{ccc}
e^\parallel \\ *e^\parallel \\  e^\perp 
\end{array}\right) ,
\label{surfxyfrenet}
\endEQ
where
\EQs
&&
\alpha = *e^\parallel \cdot e^\parallel_x 
=\Re(u e^{i\psi}) ,\quad
\beta = e^\perp \cdot e^\parallel_x 
=\Im(u e^{i\psi}) , \\
&&
\gamma= e^\perp \cdot *e^\parallel_x 
=|q|^{-1}\Im(q_x e^{i\psi}) = -\psi_x
\endEQs
and 
\EQs
&&
\mu = *e^\parallel \cdot e^\parallel_y 
=\Re(v e^{i\psi}) =q_1\alpha+|q|_x ,\quad
\rho = e^\perp \cdot e^\parallel_y
=\Im(v e^{i\psi})=q_1\beta -|q|\psi_x , \\
&&
\sigma= e^\perp \cdot *e^\parallel_y
=|q|^{-1}\Im((q_y +iv_1 q) e^{i\psi}) = v_1-\psi_y
\endEQs
are obtained through the relations \eqref{uv}, \eqref{v1}, \eqref{q}
together with the structure equations \eqref{1ststructeq}, \eqref{2ndstructeq}.
With respect to the $x,y$ coordinates on the surface, 
the scalars $\alpha$ and $\mu$ are known as the {\it geodesic curvatures};
$\beta$ and $\rho$ as the {\it normal curvatures};
$\gamma$ and $\sigma$ as the {\it relative torsions}
\cite{Gug}.

\begin{proposition}\label{extrinsic}
For a surface $\pos(x,y)$ with a parallel framing
\eqref{xnostretch}--\eqref{Erot} adapted to the $x$ coordinate lines,
satisfying the structure equations \eqref{1ststructeq} and \eqref{2ndstructeq},
the 2nd fundamental form is given by 
\EQ\label{Pi}
\Pi = (\pos_x dx+ \pos_y dy)\cdot (e^\perp_x dx + e^\perp_y dy)
= -(\beta dx dx +2\rho dx dy +(\sigma-q_1\gamma)|q| dy dy) . 
\endEQ
The components of $\Pi$ with respect to the surface tangent frame 
$e^\parallel,*e^\parallel$ yield the extrinsic curvature scalars
\EQs
&&
k_{11} = e^\parallel\cdot D_\parallel e^\perp = -\beta , \\
&&
k_{12} = e^\parallel\cdot D_{*\parallel} e^\perp =
k_{21} = *e^\parallel\cdot D_\parallel e^\perp =-\gamma, \\
&&
k_{22} = *e^\parallel\cdot D_{*\parallel} e^\perp 
=-|q|^{-1}(\sigma -q_1\gamma) ,
\endEQs
where $D_\parallel = e^\parallel\rfloor D =D_x$
and $D_{*\parallel} = *e^\parallel\rfloor D =|q|^{-1} (D_y -q_1 D_x)$
denote the projections of the total exterior derivative $D$
on the surface in the directions tangential and orthogonal 
to the $x$-coordinate lines. 
\end{proposition}

All aspects of the extrinsic surface geometry of $\pos(x,y)$
can be determined from the extrinsic curvature matrix 
$\left(\begin{array}{cc}
k_{11} & k_{12} \\ k_{21} & k_{22}
\end{array}\right)$. 
In particular, the mean curvature of the surface \cite{Gug}
\EQ\label{meancurv}
H = (k_{11}+k_{22})/2
\endEQ
is given by the normalized trace of the extrinsic curvature matrix. 

Now, in terms of the frame vectors \eqref{ex}--\eqref{e*}, 
any surface flow \eqref{tr}--\eqref{heq}
in which the $x$-coordinate lines are non-stretching
can be written in the form 
\EQ\label{surfaceeq}
\pos_t = a e^\parallel +b {*e^\parallel} +c e^\perp ,\quad |\pos_x|=1
\endEQ
with 
\EQ\label{abc}
a=D_x^{-1}\Re({\bar u}h) ,\quad
b=\Re(e^{i\psi} h) ,\quad
c=\Im(e^{i\psi} h) .
\endEQ
Through Proposition~\ref{surfaceflow}, 
we then obtain a hierarchy of flows on $\pos$
corresponding to the bi-Hamiltonian 2+1 flows on $u$ 
in Theorem~\ref{2Duhierarchy}, as given by 
\EQ\label{nthabc}
b^{(n)}+ic^{(n)} = -e^{i\psi}\mathcal{R}^{n-1}(u_y) ,\quad
a^{(n)} =-D_x^{-1}\Re({\bar u}\mathcal{R}^{n-1}(u_y))
\endEQ
in terms of the recursion operator
$\mathcal{R}=i(D_x +u D_x^{-1}\Re(u\mathcal{C}))$. 
Moreover, 
these coefficients \eqref{nthabc} have a geometrical formulation 
derived from the operator identity
\EQ\label{2DRid}
e^{i\psi}\mathcal{R} e^{-i\psi}
= iD_x +\psi_x+ iu e^{i\psi} D_x^{-1}\Re(u e^{i\psi}\mathcal{C})
= iD_x -\gamma-(\beta-i\alpha)D_x^{-1}\Re((\alpha+i\beta)\mathcal{C}) 
\endEQ
combined with 
\EQ
e^{i\psi} u_y = (e^{i\psi} u)_y -i\psi_y e^{i\psi} u 
= \alpha_y+\beta\psi_y +i(\beta_y-\alpha\psi_y) .
\endEQ
This leads to the following geometric counterpart of Theorem~\ref{2DShierarchy}.
\begin{theorem}\label{2Dsurfhierarchy}
The integrable 2+1 vector models \eqref{2Dmodeleqs}
and integrable 2+1 dynamical maps \eqref{2Dmapeqs}
correspond to a hierarchy of surface flows in $\mathbb{R}^3$,
\EQ
\pos_t 
= a^{(n-1)} e^\parallel +b^{(n-1)} {*e^\parallel} +c^{(n-1)} e^\perp ,\quad 
|\pos_x|=1, \quad n=1,2,\ldots 
\label{surfaceeqs}
\endEQ
with geometric coefficients
\EQs
b^{(n)}+i c^{(n)} = 
-\mathcal{P}^{n-1}( \alpha_y+\beta\psi_y +i(\beta_y-\alpha\psi_y) ) ,\quad
a^{(n)} = D_x^{-1}(\alpha b^{(n)}+\beta c^{(n)} ) ,
\endEQs
given in terms of the recursion operator 
\EQ
\mathcal{P} =
iD_x -\gamma-(\beta-i\alpha)D_x^{-1}\Re((\alpha+i\beta)\mathcal{C}) .
\label{Pop}
\endEQ
Here $\alpha$ and $\beta$ are the geodesic curvature and normal curvature of
the non-stretching $x$-coordinate lines on the surface $\pos(t,x,y)$,
$\gamma$ is the relative torsion of these lines, 
and $\psi =-\int\gamma dx$.
\end{theorem}

{\bf Remark:}
The bottom flow in this hierarchy 
can be written in the alternative form 
\EQ
b^{(0)} = -\rho ,\quad 
c^{(0)} = \mu ,\quad
a^{(0)} = D_x^{-1}(\beta\mu-\alpha \rho) = v_1
\endEQ
by means of the relation
\EQ\label{bottomflowid}
\mathcal{P}^{-1}( \alpha_y+\beta\psi_y +i(\beta_y-\alpha\psi_y) )
= \rho-i\mu 
\endEQ
expressed in terms of the geodesic curvature $\mu$ and normal curvature $\rho$ 
of the $y$-coordinate lines on the surface $\pos(t,x,y)$. 
This relation \eqref{bottomflowid} is obtained from 
$-i e^{i\psi} v = \mathcal{P}^{-1}(e^{i\psi} u_y)$
which is a straightforward consequence of Lemma~\ref{uvHamrel}.

Geometric properties of the surface flows in the hierarchy \eqref{surfaceeqs}
can be straightforwardly derived from the results in 
Propositions~\ref{intrinsic} and~\ref{extrinsic}
combined with the explicit evolution equations for the variables $q$ and $v$
(as determined by Lemma~\ref{uvHamrel}). 
In terms of the surface flow equation \eqref{surfaceeq}--\eqref{abc}
and the operator identity \eqref{2DRid}, 
these evolution equations are given by 
\EQs
iv_t &=& w_1 v + e^{-i\psi}(D_y +i\sigma)\mathcal{P}(b+ic) ,
\\
q_t &=& a q_x -iq w_1 +e^{-i\psi}(D_y -q_1 D_x +i(\sigma -q_1\gamma))(b+ic) ,
\endEQs
with 
\EQ
w_{1x} = \Re( (\alpha-i\beta)\mathcal{P}(b+ic) ) ,\quad
q_{1x} = |q|\alpha = -\Re( (\alpha-i\beta)\mathcal{P}(\rho-i\mu) ) ,
\endEQ
from which we obtain the evolution of 
the geodesic curvature, normal curvature, and relative torsion of
the $x$-coordinate lines:
\EQs
\alpha_t &=& 
D_x(a\alpha)+ \Re(\mathcal{D}_1{}^2(b+ic)) 
+\beta\Im(\mathcal{D}_2(b+ic)) ,\quad
\\
\beta_t &=& 
D_x(a\beta)+ \Im(\mathcal{D}_1{}^2(b+ic)) 
-\alpha\Im(\mathcal{D}_2(b+ic)) ,\quad
\\
\gamma_t &=& 
D_x( a\gamma + \Im(\mathcal{D}_2(b+ic)) )
-\Re( (\beta+i\alpha)\mathcal{D}_1(b+ic) ) ,
\endEQs
where
\EQ\label{surfaceops}
\mathcal{D}_1= D_{\parallel} -ik_{12} ,\quad
\mathcal{D}_2= D_{*\parallel} -ik_{22}
\endEQ
are a pair of geometric derivative operators 
associated with the $x$-coordinate lines
on the surface (cf. Proposition~\ref{extrinsic}). 
Similarly, we find the area element of the surface has the evolution
\EQ
(dA)_t = a(dA)_x+\Re(\mathcal{D}_2(b+ic)) dA 
= \varepsilon dA +\mathcal{L}_X dA
\endEQ
which consists of an infinitesimal change due to the tangential part of
the surface flow $X=a e^\parallel + b {*e^\parallel}$ 
plus a multiplicative expansion/contraction factor $\varepsilon = 2H c$
related to the mean curvature \eqref{meancurv} of the surface 
through the normal part of the surface flow 
$\mathcal{L}_{e^\perp} dA = 2H dA$.
These developments now lead to the following geometric results. 

\begin{theorem}\label{surfacegeom}
Under each flow $n=1,2,\ldots$ in the hierarchy \eqref{surfaceeqs},
the surface $\pos(t,x,y)$ is non-stretching the $x$-coordinate direction
while stretching in all transverse directions, such that
surface area locally expands/contracts by the dynamical factor
$$ \varepsilon^{(n)} = 2H\Im(\mathcal{P}^{n-1}(\rho-i\mu)) $$
where $H$ is the mean curvature \eqref{meancurv} of the surface. 
The geodesic curvature $\alpha$, 
normal curvature $\beta$, 
and relative torsion $\gamma=-\psi_x$ 
of the $x$-coordinate lines in each surface flow \eqref{surfaceeqs}
satisfy the integrable system of evolution equations
\EQs
\alpha_t +i \beta_t &=&
D_x(a^{(n-1)}(\alpha+i\beta)) -\mathcal{D}_1{}^2\mathcal{P}^{n-1}(\rho-i\mu) 
-(\beta-i\alpha)\Im(\mathcal{D}_2\mathcal{P}^{n-1}(\rho-i\mu)) \quad
\label{2Dcurveeq}\\
\psi_t &=&
a^{(n-1)}\psi_x + a^{(n)} +\Im(\mathcal{D}_2\mathcal{P}^{n-1}(\rho-i\mu)) 
\label{2Dtorseq}
\endEQs
in terms of the recursion operator \eqref{Pop}
and the pair of geometric operators \eqref{surfaceops},
with the geodesic curvature $\mu$ and normal curvature $\rho$ of
the $y$-coordinate lines given by \eqref{bottomflowid}, 
and where 
\EQ\label{2Dconsdens}
a^{(n)}=-D_x^{-1}\Re((\alpha-i\beta)\mathcal{P}^n(\rho-i\mu))  
\endEQ
yields (modulo total $x,y$-derivatives) 
a set of non-trivial constants of motion
$\iint_C a^{(2)}dxdy =
\frac{1}{2}\iint_C \alpha\beta_y-\beta\alpha_y -\psi_y(\alpha^2+\beta^2) dxdy$,
$\iint_C a^{(3)}dxdy =
\frac{1}{2}\iint_C \alpha_x\alpha_y+\beta_x\beta_y 
+\psi_x\psi_y(\alpha^2+\beta^2) +\psi_x(\alpha_y\beta-\beta_y\alpha) 
+\psi_y(\alpha_x\beta-\beta_x\alpha) dxdy$, 
etc. 
for the system \eqref{2Dcurveeq}--\eqref{2Dtorseq}. 
(Here $C$ denotes the coordinate domain of $(x,y)$).
\end{theorem}

We conclude by pointing out that the surface flows \eqref{surfaceeqs}
in Theorem~\ref{2Dsurfhierarchy}
and the corresponding integrable systems \eqref{2Dcurveeq}--\eqref{2Dtorseq}
in Theorem~\ref{surfacegeom}
provide a geometric realization for the 2+1 vector models \eqref{2Dmodeleqs}
and 2+1 dynamical maps \eqref{2Dmapeqs} in Theorem~\ref{2DShierarchy}.

{\bf Ex. 1}:
The $+1$ surface flow is given by 
\EQ\label{Heisenbergsurface}
\pos_t = v_1 \pos_x +\rho \pos_x\wedge \hat n +\mu \hat n ,
\quad |\pos_x|=1 ,
\endEQ
with $v_{1x}=\beta\mu-\alpha\rho$, 
where $\hat n=\sqrt{|\pos_y|^2-(\pos_x\cdot\pos_y)^2}^{-1}\pos_x\wedge\pos_y$ 
denotes the surface unit-normal.
This flow is a geometric realization of 
the 2+1 Heisenberg model \eqref{2DHeisenbergmodel},
corresponding to the new integrable 2+1 generalization of
the Schr\"odinger map \eqref{2DSchrodingermap}. 

{\bf Ex. 2}:
Similarly, 
a geometric realization of the 2+1 mKdV vector model \eqref{2Dmkdvmodel}
and the corresponding mKdV map \eqref{2Dmkdvmap} is provided by 
the $+2$ surface flow
\EQ\label{mKdVsurface}
-\pos_t = \nu_0 \pos_x +(\alpha_y+\psi_y\beta) \pos_x\wedge \hat n 
+(\beta_y-\psi_y\alpha) \hat n ,
\quad |\pos_x|=1 ,
\endEQ
with $\nu_{0x}=\alpha\alpha_y+\beta\beta_y$. 

The corresponding geometric integrable systems 
on the geodesic curvature $\alpha$, normal curvature $\beta$, 
and relative torsion $\gamma=-\psi_x$ 
of the non-stretching $x$-coordinate lines in these surface flows 
\eqref{Heisenbergsurface} and \eqref{mKdVsurface} are respectively given by
\EQs
\alpha_t +i \beta_t &=&
v_1(\alpha_x+i\beta_x) -\mathcal{D}_1{}^2(\rho-i\mu) 
+(\alpha+i\beta)( \mu\beta-\rho\alpha +i\Im(\mathcal{D}_2(\rho-i\mu)) )
\\
\psi_t &=&
v_1\psi_x -\nu_0 +\Im(\mathcal{D}_2(\rho-i\mu))
\endEQs
and
\EQs
\alpha_t +i \beta_t &=&
-\nu_0(\alpha_x+i\beta_x) 
-\mathcal{D}_1{}^2(\alpha_y+\beta\psi_y +i(\beta_y-\alpha\psi_y))
\nonumber\\&&
-(\alpha+i\beta)\big( \frac{1}{2}(\alpha^2+\beta^2)_y 
-i\Im(\mathcal{D}_2(\alpha_y+\beta\psi_y +i(\beta_y-\alpha\psi_y))) \big)
\\
\psi_t &=&
-\nu_0\psi_x +\nu_2 
+\Im(\mathcal{D}_2(\alpha_y+\beta\psi_y +i(\beta_y-\alpha\psi_y)))
\endEQs
with 
$\nu_{2x}=
\alpha\beta_{xy}-\beta\alpha_{xy}-\psi_y(\alpha\alpha_x+\beta\beta_x)
-\psi_x(\alpha\alpha_y+\beta\beta_y)-\psi_{xy}(\alpha^2+\beta^2)$.

\section{Concluding remarks}

There are some directions in which to extend the geometrical correspondence 
among integrable vector models, Hamiltonian curve and surface flows,
and bi-Hamiltonian soliton equations presented in this paper. 

First, 
it would be of interest to generalize this correspondence to integrable models 
for spin vectors $S$ in Euclidean spaces $\mathbb{R}^N$ for $N\ge 3$. 
In particular, 
the Heisenberg model $S_t=S\wedge S_{xx}$, $S\cdot S=1$, 
is well known to have a natural generalization 
where the vector wedge product and dot product in $\mathbb{R}^3$ 
are replaced by a Lie bracket $[\ ,\ ]$ 
and (negative definite) Killing form $\langle\ ,\ \rangle$
of any non-abelian semisimple Lie algebra on $\mathbb{R}^N$, i.e.
$S_t=[S,S_{xx}]$, $-\langle S,S\rangle=1$. 
All of our results in this paper have a direct extension to 
such a Lie algebra setting by applying the methods of Ref.\cite{Anco08}
to non-stretching curve flows in semisimple Lie algebras 
viewed as flat Klein geometries. 
This will lead to a large class of integrable 2+1 generalizations of 
the Heisenberg model for spin vectors $S$ in semisimple Lie algebras. 

Second, 
an interesting open problem is to find a similar geometric derivation for
non-isotropic spin vector models in $\mathbb{R}^3$ as described by 
the Landau-Lifshitz equation 
$S_t=S\wedge (S_{xx} +\J S)$, $S\cdot S=1$, 
where $\J={\rm diag}(j_1,j_2,j_3)$ is a constant matrix 
which measures the deviation from isotropy. 
This equation has two compatible Hamiltonian structures \cite{BarFokPap}, 
one of which uses the same Hamiltonian operator $S\wedge$ 
that arises in the isotropic case (i.e. in the Heisenberg model). 
The second Hamiltonian operator, however, involves the anisotropy matrix $\J$,
which cannot be derived from the frame structure equations for
non-stretching curve flows in Euclidean space. 
This suggests a non-Euclidean geometric setting will be needed instead.

\begin{acknowledgments}
S.C.A. is supported by an NSERC research grant. 
\end{acknowledgments}

\appendix
\section{Proof of Theorem~\ref{geomflows}}\label{Proof}

Let $\pos(x)$ be a space curve with $x$ as the arclength, i.e. $|\pos(x)|=1$,
and let $T,N,B$ be its Frenet frame \eqref{TNB}. 
To prove Theorem~\ref{geomflows}, 
we will enumerate the Euclidean invariants of $\pos(x)$.

Firstly, 
since the curvature $\kappa=T_x\cdot N=|T_x|$ 
and torsion $\tau=N_x\cdot B=|T_x|^{-2}T_{xx}\cdot (T\wedge T_x)$ 
are invariantly defined in terms of the unit tangent vector $T=\pos_x$ 
along $\pos(x)$, so are all of their $x$ derivatives.
This establishes part (iii) of the theorem.

Secondly, 
these (differential) invariants generate all possible scalar expressions 
formed out of $T,T_x,T_{xx},\ldots$ 
by dot products and wedge products, 
as shown from a recursive application of 
the Frenet equations \eqref{xfrenet}--\eqref{K}. 
Specifically, 
\EQs
&&
T_x=\kappa N,
\\
&&
T_{xx}=-\kappa^2T+\kappa_xN+\kappa\tau B,
\\
&&
T_{xxx}=
-3\kappa\kappa_xT+(\kappa_{xx}-\kappa^3-\kappa\tau^2)N
+(2\kappa_x\tau+\tau_x\kappa)B, 
\\&&\nonumber
\text{etc.}
\endEQs
yields
\EQs
&&
T_x\cdot T_x=-T_{xx}\cdot T=\kappa^2,
\\
&&
T_{xx}\cdot T_x=\frac{1}{2}(T_x\cdot T_x)_x=-\frac{1}{3}T_{xxx}\cdot T
=\kappa\kappa_x,
\\
&&
T_{xxx}\cdot T_x=\frac{1}{2}(T_x\cdot T_x)_{xx}-T_{xx}\cdot T_{xx}
=-\frac{1}{4}T_{xxxx}\cdot T=
\kappa\kappa_{xx}-\kappa^4-\kappa^2\tau^2,
\\&&\nonumber
\text{etc.}
\endEQs
and
\EQs
&&
T_{xx}\cdot (T\wedge T_x)=\kappa^2\tau,
\\
&&
T_{xxx}\cdot (T\wedge T_x)=(T_{xx}\cdot (T\wedge T_x))_x=
\kappa^2\tau_x+2\tau \kappa\kappa_x,
\\
&&
T_{xxx}\cdot (T\wedge T_{xx}) =
-\kappa\tau (\kappa_{xx}-\kappa^3-\kappa\tau^2)+\kappa_x(2\kappa_x\tau
+\tau_x\kappa) ,
\\&&\nonumber
\text{etc.}
\endEQs
which thus establishes parts (i) and (iv) of the theorem.

Finally, on the other hand,
since a parallel frame along $\pos(x)$ is unique 
up to a rigid ($x$-independent) rotation on the normal vectors, 
the corresponding components of the principal normal 
$T_x$ of $\pos(x)$ given by 
$u=T_x\cdot E^\perp=\kappa e^{i\theta}$ 
are invariantly defined only up to rotations 
$\theta\rightarrow\theta+\phi$, with $\phi=$const., 
on $E^\perp=(N+iB)e^{i\theta}$ 
(and likewise for the components of $T_{xx}$, $T_{xxx}$, etc.).
Such $U(1)$ rotations comprise all transformations preserving 
the parallel property \eqref{E1x}--\eqref{E23x} of this framing. 
Consequently, the actual invariants of $\pos(x)$ 
will correspond to $U(1)$-invariants formed out of 
$u$, $\bar u$, $u_x$, $\bar u_x$, $\ldots$ via the relations 
$\kappa=|u|$, $\tau=(\arg u)_{x}$. 
This establishes part (ii) of the theorem.

\section{Hamiltonian structure of the 1+1 and 2+1 Schr\"odinger maps}
\label{SchrodingerHamProof}

We will first verify the second Hamiltonian 
for the Schr\"odinger map equation \eqref{Schrodinger2ndHam}
and its 2+1 generalization \eqref{2DSchrodingerHam}. 
The following preliminaries concerning 
the tangent space structure of $\rm S^2$
will be needed. 
Here $u,v,w$ will be any triple of tangent vectors. 

(1) 
The metric tensor $g$, complex structure tensor $J$, and metric-normalized volume form $\epsilon_g$ 
on $\rm S^2$ satisfy the identities
\EQs
&
g(u,Jv)=\epsilon_g(u,v) ,
\label{Jid}\\
&
\epsilon_g(u,v)w + \epsilon_g(v,w)u + \epsilon_g(w,u)v =0 . 
\label{volid}
\endEQs
(2)
In local coordinates on $\rm S^2$, 
the metric-compatible covariant derivative $\nabla$ 
(i.e. Riemannian connection) 
and the associated covariant divergence operator $\div_g$
are given by 
\EQs
&&
\nabla_v u = \partial_v u + \Gamma_v u , 
\label{Christoffel}\\
&&
\div_g u = \div u +\tr(\Gamma_u) , 
\label{div}
\endEQs
where $\Gamma$ denotes the Christoffel symbol \cite{Gug}
determined from $g$ by the properties
\EQ\label{Christoffeleq}
\Gamma_v u = \Gamma_u v ,\quad
(\partial_w g)(u,v) = g(v,\Gamma_w u) + g(u,\Gamma_w v) . 
\endEQ
(3)
For an arbitrary variation $\delta\gamma$ of the map $\gamma$ into $\rm S^2$,
geometrically represented by a tangent vector field, 
the variation of $g|_\gamma$ and $\epsilon_g|_\gamma$
(as induced by their evaluation at $\gamma$) is given by 
\EQs
&
\delta g|_\gamma (u,v) = \partial_{\delta\gamma} g(u,v)
= g(u,\Gamma_{\delta\gamma} v) + g(v,\Gamma_{\delta\gamma} u) , 
\label{varg}\\
&
\delta \epsilon_g|_\gamma (u,v) = 
(\sqrt{\det g}^{-1}\partial_{\delta\gamma} \sqrt{\det g}) \epsilon_g (u,v)
= \tr(\Gamma_{\delta\gamma}) \epsilon_g (u,v) . 
\label{varvol}
\endEQs

Now, consider the 1+1 Hamiltonian density \eqref{2ndHam}
defined in terms of a vector field $\xi(\gamma)$ 
with covariantly-constant divergence, $\div_g\xi(\gamma)=1$. 
Through the identity \eqref{Jid}, 
this density can be written more conveniently as
\EQ
H^{(-1)} = \epsilon_g(\xi(\gamma),\gamma_x) . 
\endEQ
Its variation is given by 
\EQ\label{var-1H}
\delta H^{(-1)} = 
\tr(\Gamma_{\delta\gamma}) \epsilon_g(\xi(\gamma),\gamma_x)
+ \epsilon_g(\partial_{\delta\gamma}\xi(\gamma),\gamma_x)
+ \epsilon_g(\xi(\gamma),D_x\delta\gamma) . 
\endEQ
Integration by parts on the third term in \eqref{var-1H} yields
\EQ\label{3rdtermvarH}
\epsilon_g(\xi(\gamma),D_x\delta\gamma) = 
D_x( \epsilon_g(\xi(\gamma),\delta\gamma) )
-\epsilon_g(\partial_{\gamma_x}\xi(\gamma),\delta\gamma)
-\tr(\Gamma_{\gamma_x}) \epsilon_g(\xi(\gamma),\delta\gamma) . 
\endEQ
By combining the middle terms in \eqref{var-1H} and \eqref{3rdtermvarH}
via the identity \eqref{volid}, we get 
\EQ\label{divterms}
\epsilon_g(\partial_{\delta\gamma}\xi(\gamma),\gamma_x)
-\epsilon_g(\partial_{\gamma_x}\xi(\gamma),\delta\gamma)
= \epsilon_g(\delta\gamma,\gamma_x) \div\xi(\gamma) . 
\endEQ
Likewise, combining the first term in \eqref{var-1H} 
with the third term in \eqref{3rdtermvarH}, we obtain 
\EQ\label{Christoffelterms}
\epsilon_g(\xi(\gamma),\gamma_x)\tr(\Gamma_{\delta\gamma}) 
- \epsilon_g(\xi(\gamma),\delta\gamma)\tr(\Gamma_{\gamma_x}) 
= \epsilon_g(\delta\gamma,\gamma_x) \tr(\Gamma_{\xi(\gamma)}) . 
\endEQ
Hence, modulo total $x$-derivatives,
\eqref{divterms} and \eqref{Christoffelterms} combine to give
\[
\delta H^{(-1)} \equiv \epsilon_g(\delta\gamma,\gamma_x) \div_g\xi(\gamma)
= g(\delta\gamma,J\gamma_x)
\]
which yields the Hamiltonian derivative \eqref{2ndHamder}. 

Next, consider the 2+1 Hamiltonian densities 
\eqref{2D1stHam} and \eqref{2D2ndHam}. 
The previous derivation applies verbatim to the density \eqref{2D1stHam}, 
yielding its derivative \eqref{2D1stHamder}. 
For the second density \eqref{2D2ndHam}, 
we look at its two terms $H^{(-1)} =H_1 +H_2$ separately: 
\EQ
H_1 = \frac{1}{2}g(\gamma_x,\gamma_y) ,\quad
H_2= \frac{1}{2} v_1\epsilon_g(\xi(\gamma),\gamma_x) ,
\endEQ
with 
\EQ\label{v1x}
v_{1x} = \epsilon_g(\gamma_x,\gamma_y) .
\endEQ

First, the variation of $H_1$ is given by 
\EQs
\delta H_1 &&
= \frac{1}{2} g(\gamma_x,D_y\delta\gamma) 
+ \frac{1}{2} g(D_x\delta\gamma,\gamma_y)
+\frac{1}{2} g(\gamma_x,\Gamma_{\delta\gamma} \gamma_y)
+\frac{1}{2} g(\gamma_y,\Gamma_{\delta\gamma} \gamma_x)
\nonumber\\&&
= \frac{1}{2} g(\gamma_x,\nabla_y\delta\gamma) 
+ \frac{1}{2} g(\gamma_y,\nabla_x\delta\gamma)
\label{varH1}
\endEQs
through equations \eqref{Christoffel} and \eqref{Christoffeleq}. 
Integration by parts on these terms yields
\EQ
g(\gamma_x,\nabla_y\delta\gamma)
= D_y (g(\gamma_x,\delta\gamma)) -g(\delta\gamma,\nabla_y\gamma_x) ,\quad
g(\gamma_y,\nabla_x\delta\gamma)
= D_x (g(\gamma_y,\delta\gamma)) -g(\delta\gamma,\nabla_x\gamma_y)
\label{ibpvarH1}
\endEQ
where, on scalar expressions, 
a covariant derivative reduces to an ordinary total derivative. 
Hence, modulo total $x,y$-derivatives, 
substitution of \eqref{ibpvarH1} into \eqref{varH1} yields
\EQ\label{H1var}
\delta H_1 \equiv 
-\frac{1}{2} g(\delta\gamma,\nabla_x\gamma_y+ \nabla_y\gamma_x)
= g(\delta\gamma,-\nabla_y\gamma_x)
\endEQ
after we use the commutativity identity $\nabla_x\gamma_y=\nabla_y\gamma_x$
which is a consequence of the first property in \eqref{Christoffeleq}. 

Second, the variation of $H_2$ is given by 
\EQ\label{varH2}
\delta H_2 = 
\frac{1}{2}\epsilon_g(\xi(\gamma),\gamma_x) \delta v_1 
+\frac{1}{2} v_1 D_x(\epsilon_g(\xi(\gamma),\delta\gamma))
+\frac{1}{2} v_1 \epsilon_g(\delta\gamma,\gamma_x)\div_g\xi(\gamma)
\endEQ
where the last two terms come from 
\eqref{3rdtermvarH}--\eqref{Christoffelterms}.
To evaluate the first term in \eqref{varH2}, we start with 
\EQ\label{varv1}
\delta v_{1x} = 
\epsilon_g(D_x\delta\gamma,\gamma_y) + \epsilon_g(\gamma_x,D_y\delta\gamma)
+\tr(\Gamma_{\delta\gamma}) \epsilon_g(\gamma_x,\gamma_y)
\endEQ
and use integration by parts to expand the first two terms, giving
\EQs
&&
\epsilon_g(D_x\delta\gamma,\gamma_y) = 
D_x(\epsilon_g(\delta\gamma,\gamma_y)) -\epsilon_g(\delta\gamma,D_x\gamma_y)
-\tr(\Gamma_{\gamma_x}) \epsilon_g(\delta\gamma,\gamma_y) , 
\label{xtermvarv1}\\
&&
\epsilon_g(\gamma_x,D_y\delta\gamma) = 
D_y(\epsilon_g(\gamma_x,\delta\gamma)) -\epsilon_g(D_y\gamma_x,\delta\gamma)
-\tr(\Gamma_{\gamma_y}) \epsilon_g(\gamma_x,\delta\gamma) . 
\label{ytermvarv1}
\endEQs
Using the identity \eqref{volid}, we note the Christoffel terms 
in \eqref{varv1}, \eqref{xtermvarv1}, \eqref{ytermvarv1} 
combine to give $0$,
while the middle terms in \eqref{xtermvarv1} and \eqref{ytermvarv1} 
cancel due to 
\EQ\label{Did}
D_x\gamma_y-D_y\gamma_x=\nabla_x\gamma_y-\nabla_y\gamma_x=0 . 
\endEQ
Hence, \eqref{varv1} simplifies to a sum of total $x,y$-derivatives
\[
\delta v_{1x} = 
D_x(\epsilon_g(\delta\gamma,\gamma_y)) 
+D_y(\epsilon_g(\gamma_x,\delta\gamma)) . 
\]
As a result, the first term in the variation \eqref{varH2} becomes
\EQ\label{1sttermvarH2}
\frac{1}{2}\epsilon_g(\xi(\gamma),\gamma_x) \delta v_1 =
\frac{1}{2}\epsilon_g(\delta\gamma,\gamma_y) \epsilon_g(\xi(\gamma),\gamma_x)
+\frac{1}{2} D_x^{-1}(D_y\epsilon_g(\gamma_x,\delta\gamma)) 
\epsilon_g(\xi(\gamma),\gamma_x) . 
\endEQ
Integration by parts on the second term in \eqref{1sttermvarH2} yields
\EQ\label{2ndterm}
\frac{1}{2} D_x^{-1}D_y(\epsilon_g(\gamma_x,\delta\gamma)) 
\epsilon_g(\xi(\gamma),\gamma_x)
\equiv
\frac{1}{2} \epsilon_g(\gamma_x,\delta\gamma) 
D_x^{-1} D_y(\epsilon_g(\xi(\gamma),\gamma_x)) . 
\endEQ
Now we use the relation
\EQs
D_y \epsilon_g(\xi(\gamma),\gamma_x) - D_x \epsilon_g(\xi(\gamma),\gamma_y)  
&= &
\epsilon_g(\partial_{\gamma_y}\xi(\gamma),\gamma_x) 
- \epsilon_g(\partial_{\gamma_x}\xi(\gamma),\gamma_y) 
+ \epsilon_g(\xi(\gamma),D_y\gamma_x - D_x\gamma_y)
\nonumber\\&& 
+\epsilon_g(\xi(\gamma),\gamma_x)\tr(\Gamma_{\gamma_y})
- \epsilon_g(\xi(\gamma),\gamma_y)\tr(\Gamma_{\gamma_x})
\nonumber\\
&=& 
\epsilon_g(\gamma_y,\gamma_x)( \div\xi(\gamma) +\tr(\Gamma_{\xi(\gamma)}) )
= -v_{1x} \div_g\xi(\gamma)
\nonumber
\endEQs
obtained via the identities \eqref{volid} and \eqref{Did}. 
Thus, \eqref{2ndterm} reduces to 
\EQ
\frac{1}{2} D_x^{-1}D_y(\epsilon_g(\gamma_x,\delta\gamma)) 
\epsilon_g(\xi(\gamma),\gamma_x)
\equiv
\frac{1}{2} \epsilon_g(\gamma_x,\delta\gamma) \epsilon_g(\xi(\gamma),\gamma_y)
+\frac{1}{2} v_1 \epsilon_g(\delta\gamma,\gamma_x)
\endEQ
which combines with the first term in \eqref{1sttermvarH2}
by use of \eqref{volid} to give
\EQ\label{termsvarH2}
\frac{1}{2}\epsilon_g(\xi(\gamma),\gamma_x) \delta v_1 
\equiv 
\frac{1}{2}\epsilon_g(\xi(\gamma),\delta\gamma) \epsilon_g(\gamma_x,\gamma_y)
+\frac{1}{2} v_1 \epsilon_g(\delta\gamma,\gamma_x) . 
\endEQ
Substituting \eqref{termsvarH2} into the variation \eqref{varH2},
and using \eqref{v1x}, we get
\EQ\label{H2var}
\delta H_2 \equiv
\frac{1}{2} v_{1x} \epsilon_g(\xi(\gamma),\delta\gamma) 
+\frac{1}{2} v_1 D_x(\epsilon_g(\xi(\gamma),\delta\gamma))
+ v_1 \epsilon_g(\delta\gamma,\gamma_x)
\equiv g(\delta\gamma,v_1 J\gamma_x) . 
\endEQ

Finally, we combine the separate variations \eqref{H2var} and \eqref{H1var} 
to obtain
\EQ
\delta H^{(-1)} = \delta H_1 + \delta H_2 \equiv
g(\delta\gamma,-\nabla_y\gamma_x + v_1 J\gamma_x)
\endEQ
which yields the Hamiltonian derivative \eqref{2D2ndHamder}.

\section{Hamiltonian structure of the 1+1 and 2+1 Heisenberg models}
\label{HeisenbergHamProof}

We will next verify the second Hamiltonian 
for the 1+1 Heisenberg model \eqref{Heisenberg2ndHam}, 
given by the density
\EQ\label{-1H}
H^{(-1)} = \xi(S)\cdot(S\wedge S_x) . 
\endEQ
Here $\xi(S)$ is a vector function, defined in terms of the spin vector $S$,
such that 
\EQ\label{SperpXi}
S\cdot \xi(S)=0
\endEQ
and 
\EQ
\partial^\perp_S\cdot \xi(S)=1 ,
\endEQ
where $\partial^\perp_S=\partial_S -S(S\cdot\partial_S)$ is 
a component-wise gradient operator satisfying the properties \eqref{gradS}. 
We note that, due to these properties, 
$\partial^\perp_S$ has a well-defined action on any function of $S$
with $S\cdot S=1$. 
To proceed, the following algebraic preliminaries will be needed. 

(1)
A variation of $S$ consists of an arbitrary vector $\delta S$ $\perp S$, i.e.
$S\cdot \delta S=0$. 

(2) The variation of $\xi(S)$ induced by $\delta S$ is given by 
\EQ\label{varXi}
\delta \xi(S) = (\delta S\cdot \partial^\perp_S) \xi(S) .
\endEQ
Similarly, the total $x$-derivative of $\xi(S)$  is given by 
\EQ\label{DXi}
D_x \xi(S) = (S_x\cdot\partial^\perp_S) \xi(S) .
\endEQ

(3)
Since the subspace of vectors orthogonal to $S$ in $\mathbb{R}^3$ 
is two-dimensional, 
$\delta S\wedge S_x$ lies in the one-dimensional perp space, 
so thus
\EQ\label{crossid}
\delta S\wedge S_x = -(\delta S\cdot(S\wedge S_x)) S .
\endEQ

Now, the variation of the density \eqref{-1H} is given by 
\EQ\label{Svar-1H}
\delta H^{(-1)} = 
\delta\xi(S)\cdot(S\wedge S_x) + \xi(S)\cdot(\delta S\wedge S_x)
+\xi(S)\cdot(S\wedge D_x\delta S) .
\endEQ
Integration by parts on the third term in \eqref{Svar-1H} yields
\EQ\label{3rdtermSvar-1H}
\xi(S)\cdot(S\wedge D_x\delta S) = 
D_x( S\cdot(\delta S\wedge \xi(S)) ) -\xi(S)\cdot(S_x\wedge\delta S)
-\delta S\cdot( D_x\xi(S)\wedge S)
\endEQ
with the middle terms in \eqref{3rdtermSvar-1H} and \eqref{Svar-1H}
each vanishing due to \eqref{crossid}. 
Hence, modulo total $x$-derivatives, \eqref{Svar-1H} reduces to 
\EQ\label{equivSvar-1H}
\delta H^{(-1)} \equiv
\delta\xi(S)\cdot(S\wedge S_x) -\delta S\cdot(D_x\xi(S)\wedge S)
= S\cdot( S_x\wedge(\delta S\cdot\partial^\perp_S)\xi(S)
- \delta S\wedge(S_x\cdot\partial^\perp_S)\xi(S) )
\endEQ
via \eqref{varXi} and \eqref{DXi}. 
To simplify the terms in \eqref{equivSvar-1H}, we first rewrite
\EQ\label{mainstep}
S_x(\delta S\cdot\partial^\perp_S)
- \delta S(S_x\cdot\partial^\perp_S)
=(\delta S\wedge S_x)\wedge \partial^\perp_S
= -\delta S\cdot(S\wedge S_x) (S\wedge \partial^\perp_S)
\endEQ
by means of standard vector cross-product identities 
in addition to identity \eqref{crossid}. 
Thus, \eqref{equivSvar-1H} becomes
\EQ\label{simpSvar-1H}
\delta H^{(-1)} \equiv
(\delta S\cdot(S\wedge S_x)) S\cdot(-(S\wedge \partial^\perp_S)\wedge\xi(S))
\endEQ
and we again apply vector cross-product identities to rewrite the term
\EQ\label{doubleprodid}
(S\wedge \partial^\perp_S)\wedge\xi(S) = 
\partial^\perp_S(S\cdot\xi(S)) -\xi(S)(\partial^\perp_S\cdot S)
-S(\partial^\perp_S\cdot\xi(S)) .
\endEQ
Then, since $S\cdot\partial^\perp_S=0=S\cdot\xi(S)$, we have
\EQ\label{divXiid}
S\cdot(-(S\wedge \partial^\perp_S)\wedge\xi(S)) 
= \partial^\perp_S\cdot\xi(S) =1
\endEQ
whence \eqref{simpSvar-1H} simplifies to 
\[
\delta H^{(-1)} \equiv
\delta S\cdot(S\wedge S_x) 
\]
yielding the Hamiltonian derivative \eqref{2ndHamS}. 

The above derivation carries over verbatim to also verify 
the first Hamiltonian \eqref{2D1stHam} 
for the 2+1 Heisenberg model \eqref{2DHeisenbergHam}. 
To verify the second Hamiltonian \eqref{2D2ndHam}, 
we will separately consider the two terms in the density 
$H^{(-1)} =H_1 +H_2$ given by 
\EQ
H_1 = \frac{1}{2} S_x\cdot S_y ,\quad
H_2 = \frac{1}{2} v_1 \xi(S)\cdot( S\wedge S_x)
\endEQ
with 
\EQ\label{Sv1x}
v_{1x} = S_x\cdot(S\wedge S_y) = S\cdot(S_y\wedge S_x) . 
\endEQ
The following identity will be useful:
\EQ\label{SxSyid}
S_y\wedge S_x = v_{1x} S
\endEQ
holding similarly to \eqref{crossid}. 

First, the variation of $H_1$ is simply
\EQ\label{SH1var}
\delta H_1 = 
\frac{1}{2} S_x\cdot D_y\delta S + \frac{1}{2} S_y\cdot D_x\delta S 
\equiv -\delta S\cdot S_{xy}
\endEQ
modulo total $x,y$-derivatives. 
Next, the variation of $H_2$ consists of the terms
\EQ\label{SvarH2}
\delta H_2 = 
\frac{1}{2} \xi(S)\cdot( S\wedge S_x)\delta v_1 
+ \frac{1}{2} v_1 D_x( \delta S\cdot(\xi(S)\wedge S) )
+\frac{1}{2} v_1 \delta S\cdot(S\wedge S_x)
\endEQ
as obtained from 
\eqref{3rdtermSvar-1H}, \eqref{mainstep}, \eqref{doubleprodid}, \eqref{divXiid}.
To evaluate the first term in \eqref{SvarH2}, 
we note the variation of \eqref{Sv1x} is given by 
\EQs
\delta v_{1x} 
&& = 
\delta S\cdot(S_y\wedge S_x) 
+ S\cdot(S_y\wedge D_x\delta S) + S\cdot(D_y\delta S\wedge S_x) 
\nonumber\\
&& \equiv 
D_y( \delta S\cdot(S_x\wedge S) ) + D_x( \delta S\cdot(S\wedge S_y) )
+ 2 v_{1x} S\cdot\delta S
\nonumber
\endEQs
through \eqref{SxSyid}. 
Since the last term vanishes due to the orthogonality $\delta S$ $\perp S$,
this yields
\[
\delta v_1 = 
\delta S\cdot(S\wedge S_y) -D_x^{-1}D_y(\delta S\cdot(S\wedge S_x)) . 
\]
Hence, the first term in \eqref{SvarH2} becomes
\EQ\label{1sttermSvarH2}
\frac{1}{2} \xi(S)\cdot( S\wedge S_x)\delta v_1 
\equiv 
\frac{1}{2} \delta S\cdot(S\wedge S_y) \xi(S)\cdot(S\wedge S_x)
- \frac{1}{2} \delta S\cdot(S\wedge S_x) D_x^{-1}D_y(\xi(S)\cdot(S\wedge S_x))
\endEQ
after integration by parts. 
We simplify the second term in \eqref{1sttermSvarH2} 
by using the relations
\EQs
&&
D_y(\xi(S)\cdot(S\wedge S_x)) - D_x(\xi(S)\cdot(S\wedge S_y)) 
\nonumber\\&&
= 
S\cdot( S_x\wedge(S_y\cdot\partial^\perp_S)\xi(S)
- S_y\wedge(S_x\cdot\partial^\perp_S)\xi(S) ) 
+ 2\xi(S)\cdot(S_y\wedge S_x)
\nonumber
\endEQs
where, similarly to \eqref{mainstep} and \eqref{doubleprodid}, 
\[
S_x (S_y\cdot\partial^\perp_S) - S_y (S_x\cdot\partial^\perp_S) 
= S\cdot(S_y\wedge S_x)(S\wedge \partial^\perp_S)
\]
yields
\[
S\cdot( S_x\wedge(S_y\cdot\partial^\perp_S)\xi(S)
- S_y\wedge(S_x\cdot\partial^\perp_S)\xi(S) ) 
= -v_{1x} \partial^\perp_S\cdot\xi(S) , 
\]
while 
\[
\xi(S)\cdot(S_y\wedge S_x) =0
\]
holds due to \eqref{SxSyid} and \eqref{SperpXi}. 
Thus, we have
\[
D_x^{-1}D_y(\xi(S)\cdot(S\wedge S_x))
= \xi(S)\cdot(S\wedge S_y) -v_1 , 
\]
whence \eqref{1sttermSvarH2} simplifies to
\EQs
\frac{1}{2} \xi(S)\cdot( S\wedge S_x)\delta v_1 
&\equiv&
\frac{1}{2} \delta S\cdot(S\wedge S_y)\ \xi(S)\cdot(S\wedge S_x)
-\frac{1}{2} \delta S\cdot(S\wedge S_x)\ \xi(S)\cdot(S\wedge S_y)
\nonumber\\&&
+\frac{1}{2} v_1 \delta S\cdot(S\wedge S_x) .
\label{simp1sttermSvarH2}
\endEQs
By applying vector cross-product identities 
to the first two terms in \eqref{simp1sttermSvarH2}, we get
\EQs
&&
\frac{1}{2} (\delta S\wedge S)\cdot S_y (\xi(S)\wedge S)\cdot S_x 
- \frac{1}{2} (\delta S\wedge S)\cdot S_x (\xi(S)\wedge S)\cdot S_y
\nonumber\\&&
= \frac{1}{2} (\delta S\wedge S)\cdot( (S_x\wedge S_y)\wedge(\xi(S)\wedge S) )
= -\frac{1}{2} (\delta S\wedge S)\cdot \xi(S) v_{1x}
\nonumber
\endEQs
via \eqref{SxSyid}. 
Hence, \eqref{simp1sttermSvarH2} reduces to 
\EQ
\frac{1}{2} \xi(S)\cdot( S\wedge S_x)\delta v_1 
\equiv
\frac{1}{2} \delta S\cdot(\xi(S)\wedge S) v_{1x}
+\frac{1}{2} v_1 \delta S\cdot(S\wedge S_x) . 
\label{v1termSvarH2}
\endEQ
Finally, 
combining \eqref{v1termSvarH2} with the middle term in \eqref{SvarH2},
we get a total $x$-derivative, so thus \eqref{SvarH2} becomes
\EQ\label{SH2var}
\delta H_2 \equiv  \delta S\cdot(v_1 S\wedge S_x) . 
\endEQ

The separate variations \eqref{SH2var} and \eqref{SH1var} then combine
to yield the Hamiltonian derivative \eqref{2D2ndHamder}.

\end{document}